\documentclass[aps,prd,onecolumn,showpacs,showkeys,amsmath,amssymb]{revtex4}
\usepackage{epsfig} 
\usepackage{amsmath}  
\usepackage{makecell}
\usepackage{graphicx}
\usepackage{array}
\usepackage{subfigure}
\usepackage{bm}
\usepackage{longtable}
\usepackage{tabularx}
\usepackage{booktabs}
\usepackage{appendix}
\usepackage{caption2}
\usepackage{amsfonts}
\usepackage{amsmath}
\usepackage{graphicx}
\usepackage{subfigure}
\usepackage{dcolumn}
\usepackage{siunitx}
\usepackage{bm}
\usepackage{booktabs}
\usepackage[utf8]{inputenc}

\graphicspath{{figs/}}
\usepackage{float}
\usepackage{longtable,lscape}
\usepackage{txfonts}
\usepackage{overpic}
\usepackage{amssymb}
\usepackage{indentfirst}
\usepackage{epsfig}
\usepackage{feynmf}   
\usepackage{epstopdf}   
\usepackage{slashed}  
\usepackage{multirow}

\usepackage[colorlinks, citecolor=blue,anchorcolor=red,menucolor=red, linkcolor=red,filecolor=red,runcolor=red,urlcolor=blue,frenchlinks=red]{hyperref}

\usepackage{ulem}
\usepackage[usenames,dvipsnames]{color}

\makeatletter
\newcommand{\figcaption}{\def\@captype{figure}\caption}
\newcommand{\tabcaption}{\def\@captype{table}\caption}

\newcommand{\Rmnum}[1]{\expandafter\@slowromancap\romannumeral #1@}
\def\hlinewd#1{%
  \noalign{\ifnum0=`}\fi\hrule \@height #1 \futurelet
   \reserved@a\@xhline}
\makeatother

\begin{document}
\title{Investigation of the stability for fully-heavy $bc\bar{b}\bar{c}$ tetraquark states}

\author{Zi-Yan Yang$^1$}
\author{Qi-Nan Wang$^1$}
\author{Wei Chen$^1$}
\email{chenwei29@mail.sysu.edu.cn}
\affiliation{$^1$School of Physics, Sun Yat-sen University, Guangzhou 510275, China}
\author{Hua-Xing Chen$^2$}
\email{hxchen@seu.edu.cn}
\affiliation{$^2$School of Physics, Southeast University, Nanjing 210094, China}

\begin{abstract}
We study the existence of fully-heavy hidden-flavor $bc\bar{b}\bar{c}$ tetraquark states with various $J^{PC}=0^{\pm+}, 0^{--},1^{\pm\pm}, 2^{++}$, by using the moment QCD sum rule method augmented by fundamental inequalities. Using the moment sum rule analyses, our calculation shows that the masses for the S-wave positive parity $bc\bar{b}\bar{c}$ tetraquark states are about $12.2-12.4$ GeV in both $[\mathbf{\bar{3}_c}]_{bc}\otimes[\mathbf{3_c}]_{\bar{b}\bar{c}}$ and $[\mathbf{6_c}]_{bc}\otimes[\mathbf{\bar{6}_c}]_{\bar{b}\bar{c}}$ color configuration channels. Except for two $0^{++}$ states, such results are below the thresholds $T_{\eta_c\eta_b}/T_{\Upsilon\psi}$ and $T_{B_cB_c}$, implying that these S-wave positive parity $bc\bar{b}\bar{c}$ tetraquark states are probably stable against the strong interaction. For the P-wave negative parity $bc\bar{b}\bar{c}$ tetraquarks, their masses in the $[\mathbf{\bar{3}_c}]_{bc}\otimes[\mathbf{3_c}]_{\bar{b}\bar{c}}$ channel are around $12.9-13.2$ GeV, while a bit higher in the $[\mathbf{6_c}]_{bc}\otimes[\mathbf{\bar{6}_c}]_{\bar{b}\bar{c}}$ channel. They can decay into the $c\bar c+b\bar b$ and $c\bar b+b\bar c$ final states via the spontaneous dissociation mechanism, including the $J/\psi\Upsilon$, $\eta_c\Upsilon$, $J/\psi\eta_b$, $B_c^+B_c^-$ channels. 
\end{abstract}

\maketitle

\pagenumbering{arabic}
\section{Introduction}\label{Intro}
Recently, the LHCb Collaboration announced the evidence for a structure in the di-$J/\psi$ mass spectrum, in which a narrow resonance around 6.9 GeV  was observed with a global significance of more than $5\sigma$~\cite{LHCb2020}. They measured the mass and decay width for this $X(6900)$ state as 
$m=6905 \pm 11 \pm 7$ MeV and $\Gamma=80 \pm 19 \pm 33$ MeV assuming no interference fit, while $m=6886 \pm 11 \pm 11$ MeV and $\Gamma=168 \pm 33 \pm 69$ MeV based on an interference model. In addition, they also observed a broad structure ranging $6.2-6.8$ GeV. 
Observed in the di-$J/\psi$ mass spectrum, these structures could originate from hadron states consisting of four charm quarks~\cite{LHCb2020}.
In 2017, the CMS Collaboration had reported their measurement of an exotic structure in four lepton channel around $18.4$ GeV with a global significance of $3.6\sigma$~\cite{Khachatryan2017}, which was probably a fully-bottom tetraquark state. However, such a structure was not confirmed 
by the LHCb and latter CMS experiments~\cite{Aaij2018,Sirunyan2020}.

Before the above experimental observations, there were already some theoretical studies on the fully-heavy tetraquark systems~\cite{1975-Iwasaki-p492-492,1981-Chao-p317-317,1982-Ader-p2370-2370,1983-Ballot-p449-451,1985-Heller-p755-755,1992-Silvestre-Brac-p2179-2189,2004-Lloyd-p14009-14009,2006-Barnea-p54004-54004,2011-Berezhnoy-p94023-94023,2012-Berezhnoy-p34004-34004,Karliner2017,Wang2017,Richard2017,Chen2017}, which were much less appealing comparing with the problems of XYZ states~\cite{2016-Chen-p1-121,2017-Ali-p123-198,2017-Lebed-p143-194,2018-Guo-p15004-15004,2019-Liu-p237-320,2020-Brambilla-p1-154}. However, the research 
interest on the fully-heavy tetraquark states has been extremely inspired by the CMS and LHCb experiments, after which lots of theoretical investigations have been done based on various methods~\cite{Anwar2018,Esposito2018,Hughes2018,Wu2018,Bai2019,Chen2019b,Debastiani2019,Li2019,Liu2019,SilvestreBrac1992,Wang2019,2020-Chen-p1994-2000,Jin2020,Feng2020,Gordillo2020,Karliner2020,Liu2020,Lue2020,Ma2020,Weng2020,Yang2020,Zhao2020,Deng2020,Faustov2020,Bedolla2020}. 
In the framework of QCD sum rules~\cite{Chen2017}, we have predicted the mass spectra of the $cc\bar c\bar c$ 
and $bb\bar b\bar b$ tetraquark states with various quantum numbers. Our calculations showed that the $X(6900)$ can be interpreted as a P-wave $cc\bar c\bar c$ tetraquark state with $J^{PC}=0^{-+}$ or $1^{-+}$, while the broad resonance structure 
around $6.2-6.8$ GeV can be considered as a S-wave one with $J^{PC}=0^{++}$ or $2^{++}$. These results were also supported by our recent 
investigation on the strong decays of fully-charm tetraquarks into di-charmonia, in which we have calculated their relative branching ratios through the Fierz rearrangement~\cite{2020-Chen-p1994-2000}. 

In this work, we shall further study the hidden-flavor fully-heavy $bc\bar{b}\bar{c}$ tetraquark states in the framework of moment sum rules. Such $bc\bar{b}\bar{c}$ tetraquark systems have been investigated in the approach of chromomagnetic interaction (CMI) model~\cite{Wu2018}, in which 
the authors found that the $bc\bar{b}\bar{c}$ tetraquarks might be relatively stable and thus more favorable than the $bb\bar{b}\bar{b}$ and $cc\bar{c}\bar{c}$ states, which is also supported by Ref.~\cite{2017-Richard-p54019-54019}. In Ref.\cite{Anwar2018}, an upper bound on the mass of $bc\bar{b}\bar{c}$ tetraquark state was predicted as 12.62 GeV in a nonrelativistic effective field theory framework. Such an upper bound on mass is very close to the $J/\psi\Upsilon$ threshold, indicating that the $bc\bar{b}\bar{c}$ tetraquark state is stable against the strong interaction. However, the existence of the $bc\bar{b}\bar{c}$ tetraquark state was not supported by the constituent quark model investigation~\cite{Czarnecki2018}. More efforts are still needed to understand these controversial results. 

This paper is organized as follows. In Sec.\ref{Current}, we construct the local $bc\bar{b}\bar{c}$ tetraquark interpolating currents with various quantum numbers $J^{PC}=0^{\pm+}, 0^{--},1^{\pm\pm}, 2^{++}$. In Sec.\ref{QCDSR} we briefly introduce the formalism of QCD sum rules and moment method. We perform moment sum rule analyses for all channels and predicted the mass spectra for the $bc\bar{b}\bar{c}$ tetraquark states in Sec.\ref{Numerical}. The last section is a conclusion and discussion. 

\section{Interpolating Current for $bc\bar{b}\bar{c}$ Tetraquark}\label{Current}
\par In this section, we use the $S$-wave diquark operators $b^T_aC\gamma_5c_b$ and $b^T_aC\gamma_\mu c_b$ to compose the $bc\bar b\bar c$ tetraquark interpolating  currents. We consider the following interpolating currents with antisymmetric $[\mathbf{\bar{3}_c}]_{bc}\otimes[\mathbf{3_c}]_{\bar{b}\bar{c}}$ color structure and various spin-parity quantum numbers
\begin{equation}\label{EqCurrent}
\begin{split}
  J_{1} &=b^T_aC\gamma_5c_b(\bar{b}_a\gamma_5C\bar{c}^T_b-\bar{b}_b\gamma_5C\bar{c}^T_a), \quad J^{PC}=0^{++},\\
  J_{2} &=b^T_aC\gamma_{\mu} c_b(\bar{b}_a\gamma^{\mu} C\bar{c}^T_b-\bar{b}_b\gamma^{\mu}C\bar{c}^T_a), \quad J^{PC}=0^{++}, \\
  J_{3\mu} &=b^T_aC\gamma_{5} c_b(\bar{b}_a\gamma_{\mu} C\bar{c}^T_b-\bar{b}_b\gamma_{\mu}C\bar{c}^T_a)+b^T_aC\gamma_{\mu} c_b(\bar{b}_a\gamma_{5} C\bar{c}^T_b-\bar{b}_b\gamma_{5}C\bar{c}^T_a), \quad J^{PC}=0^{-+},1^{++}, \\
  J_{4\mu} &=b^T_aC\gamma_{5} c_b(\bar{b}_a\gamma_{\mu} C\bar{c}^T_b-\bar{b}_b\gamma_{\mu}C\bar{c}^T_a)-b^T_aC\gamma_{\mu} c_b(\bar{b}_a\gamma_{5} C\bar{c}^T_b-\bar{b}_b\gamma_{5}C\bar{c}^T_a), \quad J^{PC}=0^{--},1^{+-}, \\
  J_{5\mu\nu} &=b^T_aC\gamma_{\mu} c_b(\bar{b}_a\gamma_{\nu} C\bar{c}^T_b-\bar{b}_b\gamma_{\nu}C\bar{c}^T_a)+b^T_aC\gamma_{\nu} c_b(\bar{b}_a\gamma_{\mu} C\bar{c}^T_b-\bar{b}_b\gamma_{\mu}C\bar{c}^T_a), \quad J^{PC}=0^{++},1^{-+},2^{++}(S),0^{++}(T), \\
  J_{6\mu\nu} &=b^T_aC\gamma_{\mu} c_b(\bar{b}_a\gamma_{\nu} C\bar{c}^T_b-\bar{b}_b\gamma_{\nu}C\bar{c}^T_a)-b^T_aC\gamma_{\nu} c_b(\bar{b}_a\gamma_{\mu} C\bar{c}^T_b-\bar{b}_b\gamma_{\mu}C\bar{c}^T_a), \quad J^{PC}=1^{--},1^{+-}(A)\, , \\
\end{split}
\end{equation}
where $b, c$ stand for the bottom and charm quark fields, the subscripts $a,b$ are the color indices, $C$ denotes the charge conjugate operator, and $T$ represents the transpose of the quark fields.
For the tensor current $J_{5\mu\nu}$ and $J_{6\mu\nu}$, we list their $J^{PC}$ assignments for the traceless symmetric part $(S)$, the antisymmetric part $(A)$ and the trace $(T)$. Similarly, the corresponding color symmetric $[\mathbf{6_c}]_{bc}\otimes[\mathbf{\bar{6}_c}]_{\bar{b}\bar{c}}$ tetraquark currents are
\begin{equation}\label{EqCurrent2}
\begin{split}
  j_{1} &=b^T_aC\gamma_5c_b(\bar{b}_a\gamma_5C\bar{c}^T_b+\bar{b}_b\gamma_5C\bar{c}^T_a), \quad J^{PC}=0^{++},\\
  j_{2} &=b^T_aC\gamma_{\mu} c_b(\bar{b}_a\gamma^{\mu} C\bar{c}^T_b+\bar{b}_b\gamma^{\mu}C\bar{c}^T_a), \quad J^{PC}=0^{++}, \\
  j_{3\mu} &=b^T_aC\gamma_{5} c_b(\bar{b}_a\gamma_{\mu} C\bar{c}^T_b+\bar{b}_b\gamma_{\mu}C\bar{c}^T_a)+b^T_aC\gamma_{\mu} c_b(\bar{b}_a\gamma_{5} C\bar{c}^T_b+\bar{b}_b\gamma_{5}C\bar{c}^T_a), \quad J^{PC}=0^{-+},1^{++}, \\
  j_{4\mu} &=b^T_aC\gamma_{5} c_b(\bar{b}_a\gamma_{\mu} C\bar{c}^T_b+\bar{b}_b\gamma_{\mu}C\bar{c}^T_a)-b^T_aC\gamma_{\mu} c_b(\bar{b}_a\gamma_{5} C\bar{c}^T_b+\bar{b}_b\gamma_{5}C\bar{c}^T_a), \quad J^{PC}=0^{--},1^{+-}, \\
  j_{5\mu\nu} &=b^T_aC\gamma_{\mu} c_b(\bar{b}_a\gamma_{\nu} C\bar{c}^T_b+\bar{b}_b\gamma_{\nu}C\bar{c}^T_a)+b^T_aC\gamma_{\nu} c_b(\bar{b}_a\gamma_{\mu} C\bar{c}^T_b+\bar{b}_b\gamma_{\mu}C\bar{c}^T_a), \quad J^{PC}=0^{++},1^{-+},2^{++}(S),0^{++}(T), \\
  j_{6\mu\nu} &=b^T_aC\gamma_{\mu} c_b(\bar{b}_a\gamma_{\nu} C\bar{c}^T_b+\bar{b}_b\gamma_{\nu}C\bar{c}^T_a)-b^T_aC\gamma_{\nu} c_b(\bar{b}_a\gamma_{\mu} C\bar{c}^T_b+\bar{b}_b\gamma_{\mu}C\bar{c}^T_a), \quad J^{PC}=1^{--},1^{+-}(A). \\
\end{split}
\end{equation}
In this work, we shall use both the $[\mathbf{\bar{3}_c}]_{bc}\otimes[\mathbf{3_c}]_{\bar{b}\bar{c}}$ and $[\mathbf{6_c}]_{bc}\otimes[\mathbf{\bar{6}_c}]_{\bar{b}\bar{c}}$ tetraquark currents to investigate the $bc\bar b\bar c$ systems in QCD sum rules. 

\section{QCD Sum Rule}\label{QCDSR}
\par In this section, we will briefly introduce the formalism of QCD sum rules, which is an extensively used method to study the hadron properties 
in the past several decades~\cite{1979-Shifman-p385-447,1985-Reinders-p1-1,2000-Colangelo-p1495-1576}. 
As a start point, the two-point correlation function for the scalar or pseudoscalar currents can be written as
\begin{equation}\label{correlation0}
\Pi(q^2)=i\int d^4x e^{iq\cdot x}\langle 0|T\left[J(x)J^{\dagger}(0)\right]|0\rangle.
\end{equation}
For the vector or axial-vector current, the two-point correlation function is
\begin{equation}\label{correlation1}
\begin{split}
\Pi_{\mu\nu}(q^2)&=i\int d^4x e^{iq\cdot x}\langle 0|T\left[J_{\mu}(x)J^{\dagger}_{\nu}(0)\right]|0\rangle\\
                            &=\Pi_1(q^2)(\frac{q_\mu q_\nu}{q^2}-g_{\mu\nu})+\Pi_0(q^2)\frac{q_\mu q_\nu}{q^2}\, ,
\end{split}
\end{equation}
There are two parts of $\Pi_{\mu\nu}(q^2)$ with different Lorentz structures because $J_\mu$ is not a conserved current. The invariant function $\Pi_1(q^2)$ is related to the spin-1 hadron state, while $\Pi_0(q^2)$ is the spin-0 polarization function. The two-point correlation functions for tensor currents can be written as
\begin{equation}\label{correlation2}
\begin{split}
\Pi_{\mu\nu,\,\rho\sigma}(q^2)&=i\int d^4x e^{iq\cdot x}\langle 0|T\left[J_{\mu\nu}(x)J^{\dagger}_{\rho\sigma}(0)\right]|0\rangle\\
                            &=\left(\eta_{\mu\rho}\eta_{\nu\sigma}+\eta_{\mu\sigma}\eta_{\nu\rho}-\frac{2}{3}\eta_{\mu\nu}\eta_{\rho\sigma}\right) \Pi_{2}\left(q^{2}\right)+\cdots \, ,
\end{split}
\end{equation}
where
\begin{equation}\
\eta_{\mu\nu}=\frac{q_{\mu} q_{\nu}}{q^{2}}-g_{\mu \nu},
\end{equation} 
and  $\Pi_{2}\left(q^{2}\right)$ is the tensor current polarization functions related to the spin-2 intermediate states, and the $``\cdots"$ represents other spin-0 and spin-1 states.

We assume the current couples to the physical state $X$ through
\begin{equation}\label{coupling}
\begin{split}
 \langle 0|J|X\rangle&=f_X,\\
 \langle 0|J_\mu|X\rangle&=f_X\epsilon^X_{\mu},\\
 \langle 0|J_{\mu\nu}|X\rangle&=f_X\epsilon^X_{\mu\nu},\\
\end{split}
\end{equation}
where $f_X$ denotes the coupling constant and $\epsilon^X_{\mu},\epsilon^X_{\mu\nu}$ are the polarization vector and tensor of $X(\epsilon^X\cdot q=0)$.
\par At the hadron level, two-point correlation function can be written as
\begin{equation}
\Pi(q^2)=\frac{1}{\pi}\int^{\infty}_{s_<}\frac{\mathrm{Im}\Pi(s)}{s-q^2-i\epsilon}ds,
\end{equation}
in which we have used the form of the dispersion relation, and $s_<$ denotes the physical threshold. The imaginary part of the correlation function is defined as the spectral function, which is usually evaluated at the hadron level by inserting intermediate hadron states $\sum_n|n\rangle\langle n|$
\begin{equation}\label{spectral}
\begin{split}
\rho(s)\equiv\frac{1}{\pi}\mathrm{Im}\Pi(s)&=\sum_n\delta(s-M^2_n)\langle 0|\eta|n\rangle\langle n|\eta^\dag|0\rangle\\
&=f^2_X\delta(s-m^2_X)+\mathrm{continuum},
\end{split}
\end{equation}
where the usual parametrization of one-pole dominance for the ground state $X$ and a continuum contribution have been adopted. The spectral density $\rho(s)$ can also be evaluated at the quark-gluon level via the operator product expansion(OPE). 
To pick out the contribution of the lowest lying resonance in \eqref{spectral}, we define moments in Euclidean region $Q^2=-q^2\geq0$
\begin{equation}\label{moments}
\begin{split}
M_n(Q_0^2)=\frac{1}{n!}\left(-\frac{d}{dQ^2}\right)^n\Pi(Q^2)|_{Q^2=Q_0^2}&=\int^\infty_{s_<}\frac{\rho(s)ds}{(s+Q_0^2)^{n+1}}\\
&=\frac{f_X^2}{(m_X^2+Q_0^2)^{n+1}}[1+\delta_n(Q_0^2)],
\end{split}
\end{equation}
in which $\delta_n(Q_0^2)$ contains the contributions of higher states and continuum. It tends to zero as $n$ goes to infinity. We consider the following ratio to eliminate $f_X$ in \eqref{moments}
\begin{equation}\label{ratio}
r(n,Q_0^2)\equiv\frac{M_n(Q_0^2)}{M_{n+1}(Q_0^2)}=(m_X^2+Q_0^2)\frac{1+\delta_n(Q_0^2)}{1+\delta_{n+1}(Q_0^2)}.
\end{equation}
One expects $\delta_n(Q_0^2)\cong\delta_{n+1}(Q_0^2)$ for sufficiently large $n$ to suppress the contributions of higher states and continuum. Then hadron mass of the lowest lying resonance $m_X$ can be extracted as
\begin{equation}\label{mass}
m_X=\sqrt{r(n,Q_0^2)-Q_0^2}\, .
\end{equation}
Using the operator production expansion (OPE) method, the two-point function can also be evaluated at the quark-gluonic level as a function of various QCD parameters. In the fully heavy tetraquark systems, we only need to calculate the perturbative term and the gluon condensate contributions to the correlation functions. To evaluate the Wilson coefficients, we adopt the propagator of heavy quark in momentum space
\begin{equation}
 i S_{Q}^{a b}(p)=\frac{i \delta^{a b}}{\hat{p}-m_{Q}}
 +\frac{i}{4} g_{s} \frac{\lambda_{a b}^{n}}{2} G_{\mu \nu}^{n} \frac{\sigma^{\mu \nu}\left(\hat{p}+m_{Q}\right)+\left(\hat{p}+m_{Q}\right) \sigma^{\mu \nu}}{12}
 +\frac{i \delta^{a b}}{12}\left\langle g_{s}^{2} G G\right\rangle m_{Q} \frac{p^{2}+m_{Q} \hat{p}}{(p^{2}-m_{Q}^{2})^{4}}\, , 
\end{equation}
where $Q$ represents the $c$ or $b$ quark. The superscripts $a, b$ denote the color indices and $\hat{p}=p^{\mu}\gamma_{\mu}$. In this work, we will evaluate the perturbative term and gluon condensate term of the correlation function, the contributions from higher non-perturbative terms are small enough to be neglected. We need to emphasize that all the correlation functions are calculated at the leading order of $\alpha_s$. However, it is well known that the $\alpha_s$ correction of the perturbative term is considered to give reliable results for the charmonium states~\cite{1985-Reinders-p1-1}. Some recent studies showed that the NLO effects are also important for the doubly heavy and triply-heavy baryon systems, which will reduce the parameters dependence 
and improve the stabilities of the mass sum rules ~\cite{Chao2019,Chao2021}. However, we shall not include the $\alpha_s$ corrections in the present calculations and retain such investigation in our future work, considering the calculations will be very difficult and complicated.

\section{Numerical Analysis}\label{Numerical}
\par \par In this section we use the following quark masses and gluon condensate in our numerical analysis\cite{Nielsen2010,Narison2018,Zyla:2020zbs}:
\begin{equation}\label{inputparameter}
\begin{split}
  &\langle g_s^2GG\rangle=(0.88\pm0.25)\;\mathrm{GeV}^4,\\
  &m_b(\overline{MS})=4.18_{-0.03}^{+0.04}\;\mathrm{GeV},\\
  &m_c(\overline{MS})=1.27^{+0.03}_{-0.04}\;\mathrm{GeV}.
\end{split}
\end{equation}
We consider the renormalization scale dependence of the charm and bottom quark masses via the leading order expressions
\begin{equation}
\begin{aligned} m_{c}(\mu) &=\overline{m}_{c}\left(\frac{\alpha_{s}(\mu)}{\alpha_{s}\left(\bar{m}_{c}\right)}\right)^{12 / 25}\, ,
 \\ m_{b}(\mu) &=\overline{m}_{b}\left(\frac{\alpha_{s}(\mu)}{\alpha_{s}\left(\bar{m}_{b}\right)}\right)^{12 / 23}\, , 
 \end{aligned}
\end{equation}
where
\begin{equation}
\alpha_{s}(\mu)=\frac{\alpha_{s}\left(M_{\tau}\right)}{1+\frac{25 \alpha_{s}\left(M_{\tau}\right)}{12 \pi} \log \left(\frac{\mu^{2}}{M_{\tau}^{2}}\right)}, \quad \alpha_{s}\left(M_{\tau}\right)=0.33
\end{equation}
is determined by evolution from the $\tau$ mass using the Particle Data Group (PDG) values. For $bc\bar{b}\bar{c}$ system, we use the renormalization scale $\mu=\frac{\bar{m}_{c}+\bar{m}_{b}}{2}=2.73 \mathrm{GeV}$ in our sum rule analysis.
\par As mentioned above, there are two parameters $n$ and $Q_{0}^{2}$ in moment sum rules. To obtain a stable sum rule, one should choose suitable parameter regions for these two parameters.
In this work, we define $\xi=Q_{0}^{2}/m_{b}^{2}$ to perform sum rule analysis. In principle the parameter $n$ and $\xi$ are interconnected from the following prospects: (a) a large enough $n$ will decrease the contributions from higher states and continuum region, but it will lead to a bad OPE convergence; (b) a large $\xi$ (or $Q_{0}^{2}$) will also lead to a bad convergence of $\delta_{n}(Q_{0}^{2})$, which make it difficult to extract the parameters of the lowest lying resonance. 
\par In the following, we shall use current $J_{1}(x)$ with $J^{PC}=0^{++}$ to illustrate the details of our numerical analysis. We show the correlation function for current $J_1$ and $j_1$ as following

{\allowdisplaybreaks
\begin{eqnarray}
\nonumber
\Pi^{pert}(Q^{2})&=& \frac{c_1}{3072\pi^{6}}\int_{0}^{1}dx\int_{0}^{1}dy\int_{0}^{1}dz
\;\log\left(F\left(m_b,m_c,Q^2\right)\right)\left\{F(m_b,m_c,Q^2)^3 \left(y z^2 m_b m_c \left(4 (x-1) x y^2 z^2-4 (y-1) (z-1) z\right)\right.\right. \\
\nonumber&&\left.\left.-12 Q^2 (x-1) x (y-1) y^3 (z-1) z^5\right)+F(m_b,m_c,Q^2)^2 \left(y z^2 \left(6 Q^4 (x-1) x (y-1) y^2 (z-1) z^3-6 m_b^2 m_c^2\right)\right.\right.\\
\nonumber&&\left.\left.+y z^2 m_b m_c \left(6 Q^2 (y-1) (z-1) z-6 Q^2 (x-1) x y^2 z^2\right)\right)+3 F(m_b,m_c,Q^2)^4 (x-1) x (y-1) y^3 (z-1) z^5\right\}\\
\nonumber\Pi^{GG}(Q^{2})&=& \frac{\langle g_{s}^{2}GG \rangle}{18432\pi^{6}}\int_{0}^{1}dx\int_{0}^{1}dy\int_{0}^{1}dz
\left\{\left[\frac{(z-1) y}{(y-1)^2}+\frac{(x-1) z}{x^2}\right] m_c m_b^3+\frac{(x-1) (z-1) \left(\left(x^3-1\right) y^3+3 y^2-3 y+1\right) Q^2 z^2}{x^2 (y-1)^2}m_b^2\right.\\
\nonumber&&\left.+\frac{\left((x-1) x z y^2+\left(-x^2+x+z-1\right) y-z+1\right) z}{(x-1) x (y-1) (z-1)}m_c^2 m_b^2+\left[\frac{(y-1) y z^2}{(z-1)^2}-\frac{x}{(x-1)^2}\right] z m_c^3 m_b\right.\\
\nonumber&&\left.+\frac{\left((x-1)^2 x^2 z^2 y^4-(x-1)^2 x^2 z y^3-(z-1)^2 y^2+2 (z-1)^2 y-(z-1)^2\right) Q^2 z^2}{(x-1) x (y-1) (z-1)}m_c m_b\right.\\
\nonumber&&\left.+\frac{(y-1) Q^2 x z^2 \left((z-1)^3+(x-1)^3 y^3 z^3\right)}{(x-1)^2 (z-1)^2} m_c^2\right\} c_1 Q^2 F(m_b,m_c,Q^2)^{-1}+\left\{\left[-\frac{2 (z-1) y}{(y-1)^2}-\frac{2 (x-1) z}{x^2}\right] m_c m_b^3\right.\\
\nonumber&&\left.+2 \left[-\frac{z y^2}{(y-1)^2}+\left(-\frac{z^3}{(z-1)^2}+\frac{z}{(y-1)^2}-\frac{1}{(y-1)^2}\right) y+\frac{z}{x-x^2}-\frac{1}{x^2y}-\frac{1}{(x-1)^2y}\right] m_c^2 m_b^2\right.\\
\nonumber&&\left.-\frac{6 Q^2 (x-1) \left(\left(x^3-1\right) y^3+3 y^2-3 y+1\right) (z-1) z^2}{x^2 (y-1)^2}m_b^2+\left[\frac{2 x z}{(x-1)^2}-\frac{2 (y-1) y z^3}{(z-1)^2}\right] m_c^3 m_b\right.\\
\nonumber&&\left.+2 Q^2 \left[-\frac{3 (x-1) x z^2 y^4}{(y-1)^2}+\frac{2 (x-1) x z^3 y^3}{(z-1)^2}+\frac{3 (x-1) x z^2 y^3}{(y-1)^2}-\frac{(x-1) x z y^3}{(y-1)^2}-\frac{3 (x-1) x z^4 y^3}{(z-1)^2}+\frac{3 (y-1) (z-1) x z}{(x-1)^2}\right.\right.\\
\nonumber&&\left.\left.-\frac{3 (y-1) (z-1) z}{(x-1)^2}-\frac{3 (y-1) (z-1) z}{x}+\frac{(y-1) (z-1)}{(x-1)^2 y}+\frac{(y-1) (z-1)}{x^2 y}\right] z m_c m_b\right.\\
\nonumber&&\left.+\frac{6 (y-1) Q^2 x z^2 \left(-(z-1)^3-(x-1)^3 y^3 z^3\right)}{(x-1)^2 (z-1)^2} m_c^2+2 \left[\frac{3 (x-1) (z-1) \left(\left(x^3-1\right) y^3+3 y^2-3 y+1\right) z}{x^2 (y-1)^2}m_b^2\right.\right.\\
\nonumber&&\left.\left.+\left(\frac{3 (x-1) x z^2 y^4}{(y-1)^2}+\frac{3 (x-1) x z^4 y^3}{(z-1)^2}+\frac{2 (x-1) x z y^3}{(y-1)^2}-\frac{3 (x-1) x z^2 y^3}{(y-1)^2}-\frac{(x-1) x z^3 y^3}{(z-1)^2}+\frac{3 (y-1) (z-1) z}{(x-1)^2}\right.\right.\right.\\
\nonumber&&\left.+\frac{3 (y-1) (z-1) z}{x}-\frac{3 x (y-1) (z-1) z}{(x-1)^2}-\frac{2 (y-1) (z-1)}{(x-1)^2 y}-\frac{2 (y-1) (z-1)}{x^2 y}\right) m_c m_b\\
\nonumber&&\left.\left.\left.+\frac{3 (y-1) x z \left((z-1)^3+(x-1)^3 y^3 z^3\right)}{(x-1)^2 (z-1)^2}m_c^2\right]\right. F(m_b,m_c,Q^2) z\Bigg\}\right.c_1\log\left(F\left(m_{b},m_{c},Q^{2}\right)\right)\\
\nonumber&&-\frac{\langle g_{s}^{2}GG \rangle}{256\pi^{6}}\int_{0}^{1}dx\int_{0}^{1}dy\int_{0}^{1}dz \frac{3c_2}{y}\log\left(F\left(m_{b},m_{c},Q^{2}\right)\right)\\
\nonumber&&\left.\Bigg\{F(m_{b},m_{c},Q^{2})\right.\\
\nonumber&&\left.\Bigg(\frac{4 z\left((x-1)^2 x^2 y^4 z^2+y^2 (z-1) \left(\left(x^2-x+1\right) z-1\right)+x y^3 z (x (-z)+x+z-1)-2 y (z-1)^2+(z-1)^2\right)}{(x-1) x (y-1) (z-1)}m_b m_c\right.\\
\nonumber&&-6 Q^2 y^2 z^3 \left((x-1) x y^2 z-y z+y+z-1\right)\Bigg)\\
\nonumber&&\left.-\frac{2 Q^2 z \left((x-1)^2 x^2 y^4 z^2+y^2 (z-1) \left(\left(x^2-x+1\right) z-1\right)+x y^3 z (x (-z)+x+z-1)-2 y (z-1)^2+(z-1)^2\right)}{(x-1) x (y-1) (z-1)}m_b m_c\right.\\
\nonumber&&\left.+2 m_b^2 m_c^2 \left[\frac{1}{(x-1) x}-\frac{y^2 z}{(y-1) (z-1)}\right]+3 F(m_{b},m_{c},Q^{2})^2 y^2 z^3 \left[(x-1) x y^2 z-y z+y+z-1\right]\right.\\
&&+Q^4 y^2 z^3 \left((x-1) x y^2 z-y z+y+z-1\right)\Bigg\},
\end{eqnarray}
}

where $F(m_{c},m_{b},Q^{2})=Q^2+\left(\frac{m_b^2}{x y z}+\frac{m_{c}^{2}}{(1-x)y z}+\frac{m_b^{2}}{(1-y)z}+\frac{m_c^2}{(1-z)}\right)$ and the coefficients $c_1=12,c_2=-1/12$ are for color antisymmetric currents $J_i$ while $c_1=24,c_2=1/12$ are for color symmetric currents $j_i$.

\begin{figure}[t!]
\centering
\includegraphics[width=10cm]{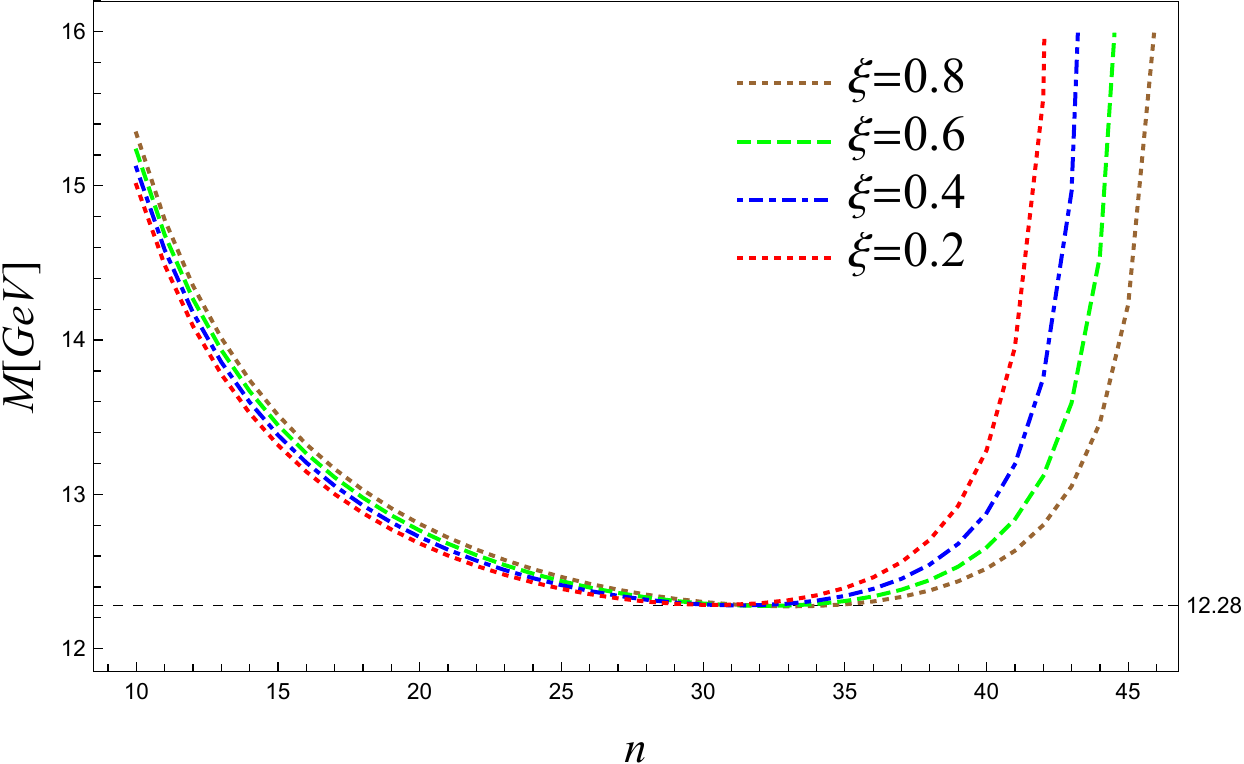}\\
\caption{Extracted hadron mass of $bc\bar{b}\bar{c}$ tetraquark with $J^{P}=0^{++}$ from $J_{1}(x)$ with respect to $n$ for different $\xi$.}
\label{massVSn}
\end{figure}
\begin{figure}[t!]
\centering
\includegraphics[width=10cm]{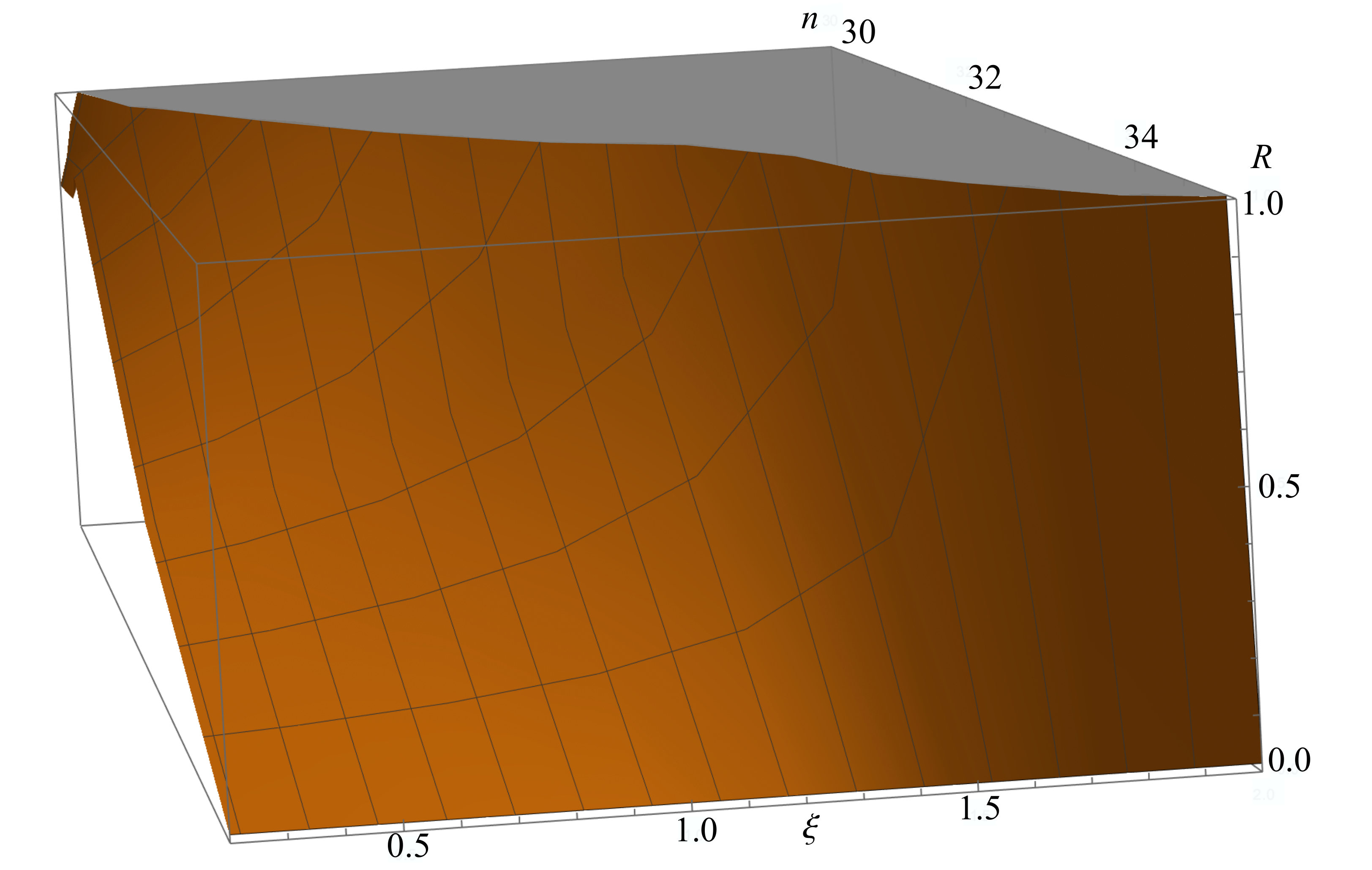}\\
\caption{Ratio $R$ as a function of $n$ and $\xi$ for $J_{1}(x)$ with $J^{P}=0^{++}$.}
\label{n_R_xi}
\end{figure}

For the parameter $n$, the upper bound can be obtained by guaranteeing the convergence of OPE series. We require that the contribution of the gluon condensate be smaller than the perturbative term, and obtain the upper bound $n_{max}=43, 44, 45, 46$ for $\xi=0.2, 0.4, 0.6, 0.8$ respectively. In Fig.~(\ref{massVSn}), we show the variation of the extracted mass with respect to $n$ for different value of $\xi$, and obtain a stable mass plateau where  the dependence of extracted mass on $n$ and $\xi$ are minimized. We find that this plateau can be obtained by studying the integral expression of the moments $M_{n}\left(Q_{0}^{2}\right)$ which satisfies the Schwarz inequality (as a special case of H{\"o}lder inequality) in the following relation
\begin{equation}
R=\frac{M_{n}\left(Q_{0}^{2}\right)^{2}}{M_{r}\left(Q_{0}^{2}\right) M_{2 n-r}\left(Q_{0}^{2}\right)} \leq 1 \, ,
\end{equation}
where $r<2n$. We show the ratio $R$ as a function of $n$ and $\xi$ in Fig.(\ref{n_R_xi}) which $R>1$ in the gray region and  $R<1$ in the yellow region.  The demarcation line between these two regions gives the values of $(n, \xi)$ for the plateaus in the mass curves. We then obtain the plateaus $(n, \xi)=(30, 0.2), (31, 0.4), (32, 0.6), (33, 0.8)$ which provide much stronger constrains for the $(n, \xi)$ plane than the OPE convergence. Then the hadron mass from the current $J_{1}(x)$ with $J^{P}=0^{++}$ is extracted as
\begin{equation}
m_{J_1}=12.28_{-0.14}^{+0.15} ~\text{GeV}\, ,
\end{equation}
where the errors come from the uncertainties of $\xi$ and $n$, the heavy quark masses and the gluon condensate.

\begin{table}[h!]
\caption{The mass spectra of $ bc\bar{b}\bar{c}$ tetraquark states with various $J^{PC}$.}\label{bcbcResultTab}
\renewcommand\arraystretch{1.3} 
\setlength{\tabcolsep}{2.em}{ 
\begin{tabular}{c c c c c}
  \hline
  \hline
 $J^{PC}$ & $Current$ & Mass$[\mathrm{GeV}]$ &   $Current$ & Mass$[\mathrm{GeV}]$ \vspace{1ex}  \\
              &    $[\mathbf{\bar{3}\otimes3}]_c$   &                                       &     $[\mathbf{6\otimes\bar{6}}]_c$     &             \vspace{1ex}  \\
  \hline
$0^{++}$ & $J_1$ & $12.28^{+0.15}_{-0.14}$ & $j_1$ & $12.37^{+0.15}_{-0.14}$ \vspace{1ex}  \\
           & $J_2$ & $12.46^{+0.17}_{-0.15}$ & $j_2$ & $12.29^{+0.15}_{-0.12}$ \vspace{1ex}  \\
           & $J_{5\mu\nu}$ & $12.35^{+0.14}_{-0.12}$ & $j_{5\mu\nu}$ & $12.32^{+0.15}_{-0.12}$ \vspace{1ex}  \\
           & $J_{5\mu\nu}(T)$ & $12.45^{+0.17}_{-0.15}$  & $j_{5\mu\nu}(T)$ & $12.29^{+0.14}_{-0.12}$ \vspace{1ex}  \\
$0^{-+}$ & $J_{3\mu}$ & $12.99^{+0.22}_{-0.18}$ & $j_{3\mu}$ & $13.16^{+0.23}_{-0.20}$ \vspace{1ex}  \\
$0^{--}$ & $J_{4\mu}$ & $12.98^{+0.22}_{-0.18}$ & $j_{4\mu}$ & $13.17^{+0.23}_{-0.19}$ \vspace{1ex}  \\        
$1^{++}$ & $J_{3\mu}$ & $12.30^{+0.15}_{-0.14}$ & $j_{3\mu}$ & $12.36^{+0.16}_{-0.14}$ \vspace{1ex}  \\
$1^{+-}$ & $J_{4\mu}$ & $12.32^{+0.15}_{-0.13}$ & $j_{4\mu}$ & $12.34^{+0.15}_{-0.14}$ \vspace{1ex}  \\
              & $J_{6\mu\nu}(A)$ & $12.38^{+0.13}_{-0.12}$ & $j_{6\mu\nu}(A)$ & $12.30^{+0.14}_{-0.12}$ \vspace{1ex}  \\
$1^{-+}$ & $J_{5\mu\nu}$ & $13.23^{+0.24}_{-0.20}$ & $j_{5\mu\nu}$ & $13.17^{+0.23}_{-0.20}$ \vspace{1ex}  \\
$1^{--}$ & $J_{6\mu\nu}$ & $12.91^{+0.19}_{-0.16}$ & $j_{6\mu\nu}$ & $13.13^{+0.22}_{-0.19}$ \vspace{1ex}  \\          
$2^{++}$ & $J_{5\mu\nu}$ & $12.30^{+0.15}_{-0.14}$ & $j_{5\mu\nu}$ & $12.35^{+0.15}_{-0.14}$ \vspace{1ex}  \\
  \hline
  \hline
\end{tabular}
}
\end{table}
By performing the same numerical analyses to the all interpolating currents from Eq.~(\ref{EqCurrent}-\ref{EqCurrent2}), we obtain the mass spectra for the $bc\bar b\bar c$ tetraquark states in various channels and collect them in Table~\ref{bcbcResultTab}. In general, a diquark-antidiquark tetraquark state should be an admixture of the two color configurations of $[\mathbf{6_c}]_{bc}\otimes[\mathbf{\bar{6}_c}]_{\bar{b}\bar{c}}$ and $[\mathbf{\bar{3}_c}]_{bc}\otimes[\mathbf{3_c}]_{\bar{b}\bar{c}}$. Such mixing effect will affect the mass spectra of the tetraquark states by the couple-channel interactions between the diquark and antidiquark fields. For the interpolating currents with $[\mathbf{\bar{3}_c}]_{bc}\otimes[\mathbf{3_c}]_{\bar{b}\bar{c}}$ antisymmetric color structure, the masses for positive parity $bc\bar{b}\bar{c}$ states are about $12.2-12.4$ GeV while $12.9-13.2$ GeV for the negative parity $bc\bar{b}\bar{c}$ states. For the color symmetric interpolating currents with $[\mathbf{6_c}]_{bc}\otimes[\mathbf{\bar{6}_c}]_{\bar{b}\bar{c}}$, it is shown that the masses for positive parity states are almost the same with those in the color antisymmetric channels, while the masses for negative parity states are slightly higher than those in the color antisymmetric channels. 
Such results are consistent with the conclusions of Refs.~\cite{Deng2020,Lue2020}, in which the Coulomb interaction plays a important role in the  
$[Q_1Q_2][\bar{Q}_3\bar{Q}_4]$ systems and leads to the mass splitting between different color configurations. 
In Ref.~\cite{Wang:2019rdo}, the couple-channel effect of these two color configurations has been studied in two nonrelativistic quark models, indicating 
that such mixing effects are induced by the hyperfine interactions between the diquark and antidiquark. 

\begin{figure}[t!]
\centering
\includegraphics[width=8cm]{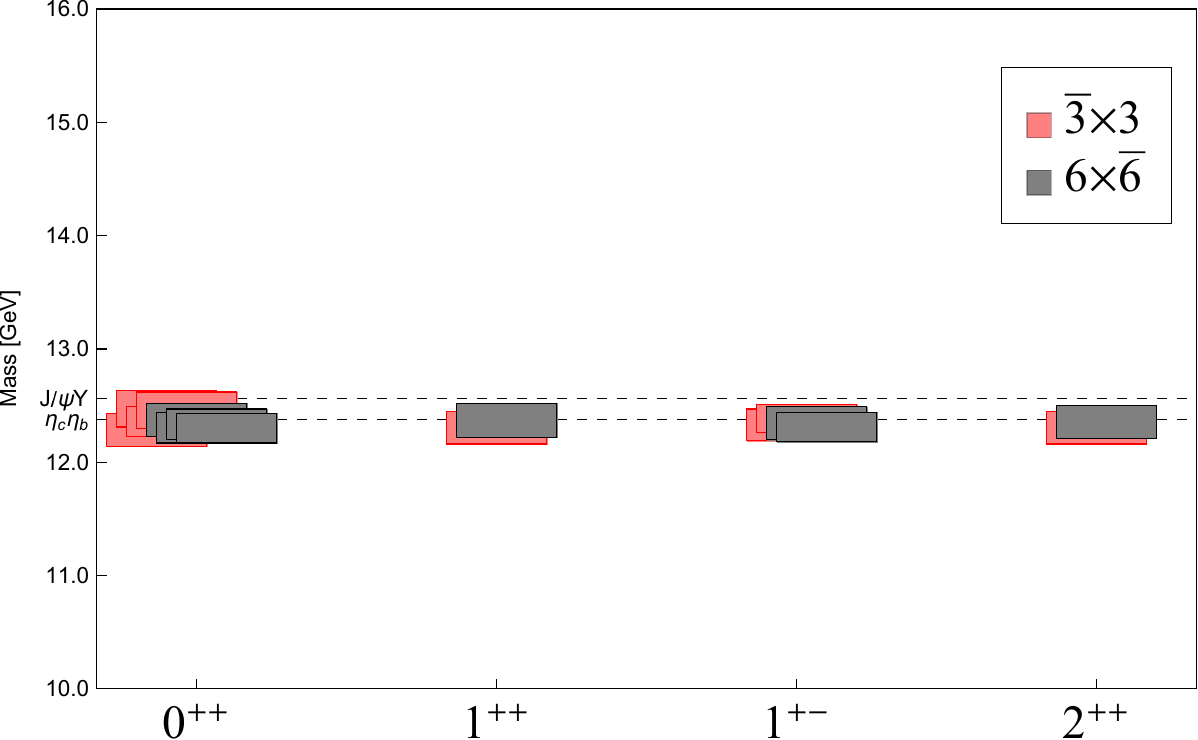}\quad
\includegraphics[width=8cm]{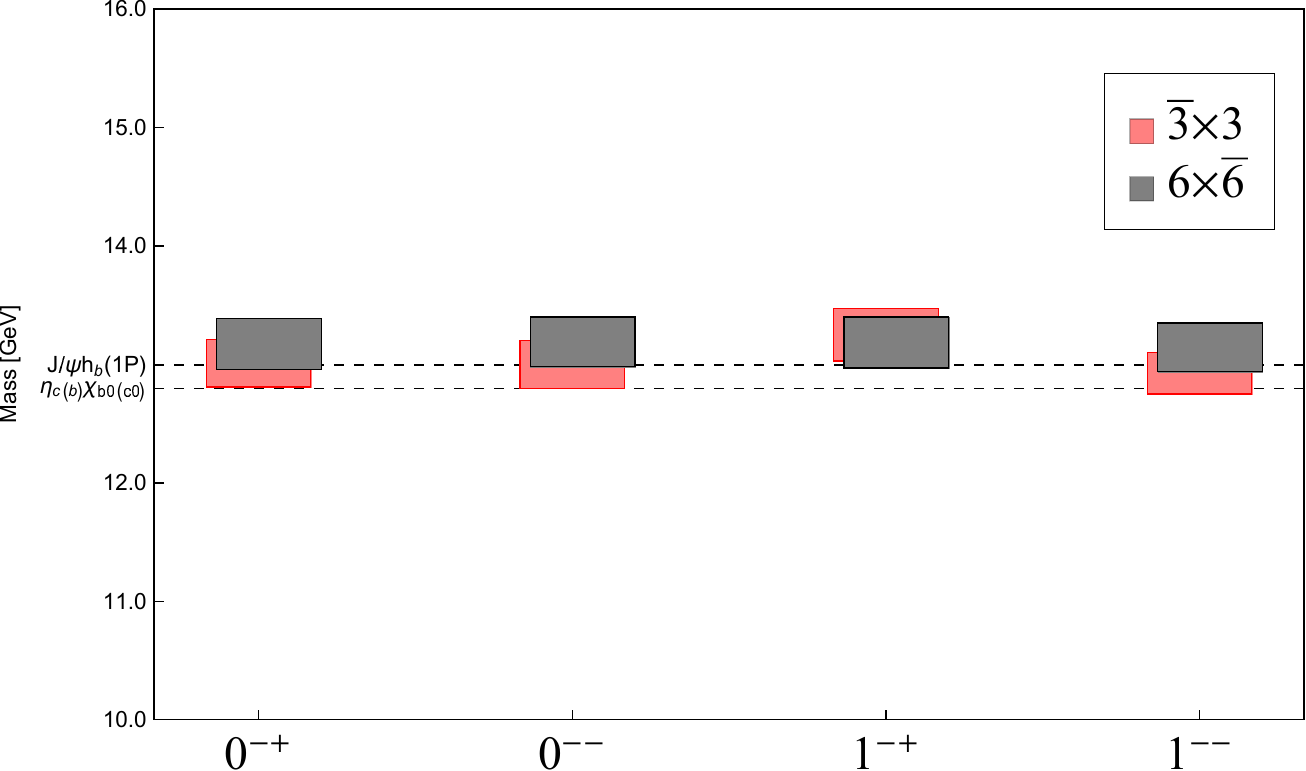}\\
\caption{Mass spectra for the $bc\bar{b}\bar{c}$ tetraquark states, comparing with the corresponding two-meson thresholds.}
\label{bcbcResultFig}
\end{figure}

\begin{table}[h!]
\caption{Possible decay modes of the $bc\bar{b}\bar{c}$ tetraquark states.}\label{bcbcDecayTab}
\renewcommand\arraystretch{1.3} 
\setlength{\tabcolsep}{2.em}{ 
\begin{tabular}{c c c }
  \hline
   \hline
 $J^{PC}$ & $S-$wave & $P-$wave \\
  \hline

  $0^{++}$ & $\eta_c\eta_b$ & $-$\vspace{1ex}\\ 
  $0^{-+}$ & $\eta_c\chi_{b0}(1P),\chi_{c0}(1P)\eta_b,$ & $J/\psi\Upsilon$\vspace{1ex}\\ 
   & $h_c(1P)\Upsilon,J/\psi h_b(1P)$ &\vspace{1ex} \\ 
  $0^{--}$  &  $\chi_{c1}(1P)\Upsilon,J/\psi \chi_{b1}(1P)$ & $\eta_c\Upsilon,J/\psi\eta_b$ \vspace{1ex}\\
  $1^{++}$ & $-$ & $-$ \vspace{1ex}\\
  $1^{+-}$ & $-$ & $-$\vspace{1ex}\\
  $1^{-+}$ & $\eta_c\chi_{b1}(1P),\chi_{c1}(1P)\eta_b,$ & $J/\psi\Upsilon,B_c^-B_c^{+}$\vspace{1ex}\\
                & $h_c(1P)\Upsilon,J/\psi h_b(1P)$ &\vspace{1ex} \\
  $1^{--}$ & $\chi_{c0}(1P)\Upsilon,J/\psi\chi_{b0}(1P),$ & $\eta_c\Upsilon,J/\psi\eta_b,$\vspace{1ex}\\
               & $\chi_{c1}(1P)\Upsilon,J/\psi \chi_{b1}(1P)$ & $B_c^-B_c^{+}$ \vspace{1ex}\\
               & $\eta_c h_{b}(1P),h_{c}(1P)\eta_b$ & \vspace{1ex}\\
  $2^{++}$ & $-$ & $-$ \vspace{1ex}\\         
  \hline
   \hline
\end{tabular}
}
\end{table}

\section{Conclusion and Discussion}\label{Resultanddis}
\par We have calculated the mass spectra for the $bc\bar{b}\bar{c}$ tetraquark states in the framework of QCD moment sum rules. We construct the interpolating tetraquark currents and calculate their two-point correlation functions containing perturbative term and gluon condensate term at the leading order of $\alpha_s$. We have performed the moment sum rule analyses to find stable sum rules for all currents and obtained the reliable mass spectra for the $bc\bar{b}\bar{c}$ tetraquark states. Our results show that the masses for the positive parity $bc\bar{b}\bar{c}$ tetraquark states are around $12.2-12.4$ GeV in both color configurations $[\mathbf{\bar{3}_c}]_{bc}\otimes[\mathbf{3_c}]_{\bar{b}\bar{c}}$ and $[\mathbf{6_c}]_{bc}\otimes[\mathbf{\bar{6}_c}]_{\bar{b}\bar{c}}$ channels. For the negative parity $bc\bar{b}\bar{c}$ systems, the masses are about $12.9-13.2$ GeV in 
the $[\mathbf{\bar{3}_c}]_{bc}\otimes[\mathbf{3_c}]_{\bar{b}\bar{c}}$ channel, while a bit higher for the $[\mathbf{6_c}]_{bc}\otimes[\mathbf{\bar{6}_c}]_{\bar{b}\bar{c}}$ channel. 

In general, the $bc\bar{b}\bar{c}$ tetraquark states can mainly decay into a charmonium plus a bottomonium final states or two $B_c$ mesons so long as the kinematics allows. There are thus two mass thresholds $T_{\eta_c\eta_b}/T_{\Upsilon\psi}$ and $T_{B_cB_c}$ for considering the strong decays of the $bc\bar{b}\bar{c}$ tetraquarks.  In Fig.~\ref{bcbcResultFig}, we show the mass spectra of $bc\bar{b}\bar{c}$ tetraquark states and compare them with the corresponding $T_{\eta_c\eta_b}/T_{\Upsilon\psi}$ thresholds, since these two thresholds are lower than $T_{B_cB_c}$. Except for two $0^{++}$ states in the $[\mathbf{\bar{3}_c}]_{bc}\otimes[\mathbf{3_c}]_{\bar{b}\bar{c}}$ structure, one finds that all S-wave positive parity $bc\bar{b}\bar{c}$ states are below or very close to the $T_{\eta_c\eta_b}$ and $T_{\Upsilon\psi}$ thresholds, implying that these states are very difficult to decay into the two-meson final states by spontaneous dissociation mechanism. These positive parity $bc\bar{b}\bar{c}$ tetraquarks are probably stable 
against the strong interaction. Comparing with our predictions for the $cc\bar c\bar c$ tetraquark states in Ref.~\cite{Chen2017}, these $bc\bar{b}\bar{c}$ tetraquarks are more stable and favorable than the four-charm states, which is consistent with the results in Refs.~\cite{2017-Richard-p54019-54019,Wu2018}. For the P-wave negative parity $bc\bar{b}\bar{c}$ tetraquark states, they lie above the corresponding mass thresholds and thus can decay via spontaneous dissociation mechanism. Considering the kinematics constraint and conservation of $J^{PC}$ quantum numbers, we list the possible two-meson strong decay channels for all $bc\bar{b}\bar{c}$ tetraquark states in Table \ref{bcbcDecayTab}, including the S-wave and P-wave decay channles. 

Our calculations provide some understanding of the stability for the $bc\bar{b}\bar{c}$ tetraquark states. If such tetraquark states exist, the positive parity $bc\bar{b}\bar{c}$ states may be stable (except for two $0^{++}$ states) and can only decay via the electromagnetic and weak interactions. However, the negative parity $bc\bar{b}\bar{c}$ states can decay into the $c\bar c+b\bar b$ and $c\bar b+b\bar c$ final states, including the $J/\psi\Upsilon$, $\eta_c\Upsilon$, $J/\psi\eta_b$, $B_c^+B_c^-$ channels. These $bc\bar{b}\bar{c}$ tetraquark states may be observed at facilities such as LHCb, CMS and RHIC in the future.

\section*{ACKNOWLEDGMENTS}
This work is supported in part by National Key R$\&$D Program of China under Contracts No. 2020YFA0406400, the National Natural Science Foundation of China under Grants No. 11722540 and No. 12075019, the Fundamental Research Funds for the Central Universities.
\section*{Appendix}
\subsection{Correlation function for interpolating currents}
\par The correlation function for current $J_2$ and $j_2$ is shown as:
{\allowdisplaybreaks
\begin{eqnarray}
\nonumber
\Pi^{pert}(Q^{2})&=& \frac{1}{768 \pi ^6}\int_{0}^{1}dx\int_{0}^{1}dy\int_{0}^{1}dz\;
c_1 F(m_b,m_c,Q^2)^2 y z^2 \log\left(F\left(m_b,m_c,Q^2\right)\right) \\
\nonumber&&\left(2 F(m_b,m_c,Q^2) (x-1) x y^2 z^2 m_b m_c-2 F(m_b,m_c,Q^2) (y-1) (z-1) z m_b m_c-3 Q^2 (x-1) x y^2 z^2 m_b m_c\right.\\
\nonumber&&+3 Q^2 (y-1) (z-1) z m_b m_c-6 m_b^2 m_c^2+3 F(m_b,m_c,Q^2)^2 (x-1) x (y-1) y^2 (z-1) z^3\\
\nonumber&&\left.-12 F(m_b,m_c,Q^2) Q^2 (x-1) x (y-1) y^2 (z-1) z^3+6 Q^4 (x-1) x (y-1) y^2 (z-1) z^3\right)\\
\nonumber\Pi^{GG}(Q^{2})&=& \frac{\langle g_{s}^{2}GG \rangle}{9216 \pi ^6}\int_{0}^{1}dx\int_{0}^{1}dy\int_{0}^{1}dz\\
\nonumber&&c_1 \left\{2 \left[-\frac{6 Q^2 (x-1) x (y-1) y^3 m_c^2 z^5}{(z-1)^2}-\frac{3 Q^2 (x-1) x y^3 m_b m_c z^5}{(z-1)^2}+\frac{2 (x-1) Q^2 x y^3 m_b m_c z^4}{(z-1)^2}+\frac{3 (x-1) Q^2 x y^3 m_b m_c z^3}{(y-1)^2}\right.\right.\\
\nonumber&&-\frac{3 Q^2 (x-1) x y^4 m_b m_c z^3}{(y-1)^2}-\frac{(y-1) y m_b m_c^3 z^3}{(z-1)^2}-\frac{2 y m_b^2 m_c^2 z^3}{(z-1)^2}+\frac{6 (x-1) (y-1) (z-1) Q^2 m_b^2 z^2}{x^2}\\
\nonumber&&+\frac{3 (y-1) (z-1) Q^2 x m_b m_c z^2}{(x-1)^2}-\frac{6 Q^2 x (y-1) (z-1) m_c^2 z^2}{(x-1)^2}-\frac{3 Q^2 (y-1) (z-1) m_b m_c z^2}{(x-1)^2}\\
\nonumber&&-\frac{6 Q^2 (x-1) x y^3 (z-1) m_b^2 z^2}{(y-1)^2}-\frac{Q^2 (x-1) x y^3 m_b m_c z^2}{(y-1)^2}-\frac{3 Q^2 (y-1) (z-1) m_b m_c z^2}{x}\\
\nonumber&&+\frac{x m_b m_c^3 z}{(x-1)^2}+\frac{2 y m_b^2 m_c^2 z}{(y-1)^2}+\frac{2 m_b^2 m_c^2 z}{(x-1)^2}+\frac{2 m_b^2 m_c^2 z}{x}+F\left(m_b,m_c,Q^2\right) z\left(\frac{3 (x-1) x z^2 m_b m_c y^4}{(y-1)^2}+\frac{6 (x-1) (z-1) x z m_b^2 y^3}{(y-1)^2}\right.\\
\nonumber&&+\frac{6 (x-1) (y-1) x z^4 m_c^2 y^3}{(z-1)^2}+\frac{3 (x-1) x z^4 m_b m_c y^3}{(z-1)^2}+\frac{2 (x-1) x z m_b m_c y^3}{(y-1)^2}-\frac{3 (x-1) x z^2 m_b m_c y^3}{(y-1)^2}\\
\nonumber&&-\frac{(x-1) x z^3 m_b m_c y^3}{(z-1)^2}+\frac{6 (y-1) (z-1) x z m_c^2}{(x-1)^2}+\frac{3 (y-1) (z-1) z m_b m_c}{(x-1)^2}+\frac{3 (y-1) (z-1) z m_b m_c}{x}\\
\nonumber&&\left.-\frac{3 x (y-1) (z-1) z m_b m_c}{(x-1)^2}-\frac{6 (x-1) (y-1) (z-1) z m_b^2}{x^2}-\frac{2 (y-1) (z-1) m_b m_c}{(x-1)^2 y}-\frac{2 (y-1) (z-1) m_b m_c}{x^2 y}\right)\\
\nonumber&&+\frac{(y-1) (z-1) Q^2 m_b m_c z}{(x-1)^2 y}+\frac{(y-1) (z-1) Q^2 m_b m_c z}{x^2 y}-\frac{2 x m_b^2 m_c^2 z}{(x-1)^2}-\frac{2 y^2 m_b^2 m_c^2 z}{(y-1)^2}-\frac{(x-1) m_b^3 m_c z}{x^2}-\frac{2 y m_b^2 m_c^2}{(y-1)^2}\\
\nonumber&&\left.-\frac{y (z-1) m_b^3 m_c}{(y-1)^2}-\frac{2 m_b^2 m_c^2}{(x-1)^2 y}-\frac{2 m_b^2 m_c^2}{x^2 y}\right] Log\left[F\left(m_b,m_c,Q^2\right)\right]+\left[\frac{2 (x-1) (y-1) Q^2 x y^3 m_c^2 z^5}{(z-1)^2}\right.\\
\nonumber&&+\frac{(x-1) Q^2 x y^3 m_b m_c z^5}{(z-1)^2}+\frac{2 y^2 m_b^2 m_c^2 z}{(y-1)^2}+\frac{2 x m_b^2 m_c^2 z}{(x-1)^2}+\frac{(x-1) m_b^3 m_c z}{x^2}-\frac{x m_b m_c^3 z}{(x-1)^2}-\frac{2 m_b^2 m_c^2 z}{(x-1)^2}-\frac{2 y m_b^2 m_c^2 z}{(y-1)^2}\\
\nonumber&&-\frac{Q^2 (x-1) x y^3 m_b m_c z^4}{(z-1)^2}+\frac{(y-1) y m_b m_c^3 z^3}{(z-1)^2}+\frac{2 y m_b^2 m_c^2 z^3}{(z-1)^2}+\frac{(x-1) Q^2 x y^4 m_b m_c z^3}{(y-1)^2}-\frac{Q^2 (x-1) x y^3 m_b m_c z^3}{(y-1)^2}\\
\nonumber&&+\frac{2 (x-1) (z-1) Q^2 x y^3 m_b^2 z^2}{(y-1)^2}+\frac{2 (y-1) (z-1) Q^2 x m_c^2 z^2}{(x-1)^2}+\frac{(y-1) (z-1) Q^2 m_b m_c z^2}{(x-1)^2}+\frac{(y-1) (z-1) Q^2 m_b m_c z^2}{x}\\
\nonumber&&-\frac{Q^2 x (y-1) (z-1) m_b m_c z^2}{(x-1)^2}-\frac{2 y m_b^2 m_c^2 z^2}{(z-1)^2}-\frac{2 Q^2 (x-1) (y-1) (z-1) m_b^2 z^2}{x^2}\\
\nonumber&&\left.\left.-\frac{2 m_b^2 m_c^2 z}{x}+\frac{(z-1) y m_b^3 m_c}{(y-1)^2}\right] Q^2F\left(m_b,m_c,Q^2\right)^{-1}\right\}\\
\nonumber&&-\frac{\langle g_{s}^{2}GG \rangle}{128 \pi ^6}\int_{0}^{1}dx\int_{0}^{1}dy\int_{0}^{1}dz\; \frac{3 c_2}{(x-1) x (y-1) (z-1)} \log\left(F\left(m_b,m_c,Q^2\right)\right) \\
\nonumber&&\left\{-z m_b m_c \left[-y^2 z \left(x^2 (2 z-1)-2 x z+x+z-1\right)+y (z-1) \left(2 x^2 z-2 x z+z+1\right)+(x-1) x y^3 z^2-z+1\right]\right.\\
\nonumber&& \left(2 F\left(m_b,m_c,Q^2\right)\right.\left.\left.-Q^2\right)-(x-1) x (y-1) y^2 (z-1) z^3 (y z-1) \left[-6 Q^2 F\left(m_b,m_c,Q^2\right)+3 F\left(m_b,m_c,Q^2\right)^2+Q^4\right]\right.\\
\nonumber&&\left.+2 m_b^2 m_c^2 (y z-1)\right\},\\
\end{eqnarray}
}
The correlation function for current $J_3$ and $j_3$ is shown as:
{\allowdisplaybreaks
\begin{eqnarray}
\nonumber
\Pi^{pert}_0(Q^{2})&=& \frac{1}{1024 \pi ^6}\int_{0}^{1}dx\int_{0}^{1}dy\int_{0}^{1}dz\;c_1 y z^2 F\left(m_b,m_c,Q^2\right)^2 \log \left(F\left(m_b,m_c,Q^2\right)\right)\\
\nonumber&&\left\{4 Q^2 (x-1) x (y-1) y^2 (z-1) z^3 F\left(m_b,m_c,Q^2\right)+(x-1) x (y-1) y^2 (z-1) z^3 F\left(m_b,m_c,Q^2\right){}^2\right.\\
\nonumber&&\left.+2 (x-1) x y^2 z^2 m_b m_c F\left(m_b,m_c,Q^2\right)-2 (y-1) (z-1) z m_b m_c F\left(m_b,m_c,Q^2\right)-4 m_b^2 m_c^2\right.\\
\nonumber&&\left.-4 Q^4 (x-1) x (y-1) y^2 (z-1) z^3\right\}
\\
\nonumber\Pi_0^{GG}(Q^{2})&=& \frac{\langle g_{s}^{2}GG \rangle}{18432 \pi ^6}\int_{0}^{1}dx\int_{0}^{1}dy\int_{0}^{1}dz\\
\nonumber&&c_1 \Bigg\{2 z\frac{1}{{F\left(m_b,m_c,Q^2\right)}} \Bigg[-\frac{Q^2 (x-1) x (y-1) y^3 m_c^2 z^4}{(z-1)^2}+\frac{(x-1) (y-1) (z-1) Q^2 m_b^2 z}{x^2}-\frac{Q^2 x (y-1) (z-1) m_c^2 z}{(x-1)^2}\\
\nonumber&&+\frac{y m_b^2 m_c^2 z^2}{(z-1)^2}-\frac{Q^2 (x-1) x y^3 (z-1) m_b^2 z}{(y-1)^2}-\frac{y m_b^2 m_c^2 z}{(z-1)^2}+\frac{y^2 m_b^2 m_c^2}{(y-1)^2}+\frac{x m_b^2 m_c^2}{(x-1)^2}-\frac{m_b^2 m_c^2}{(x-1)^2}-\frac{y m_b^2 m_c^2}{(y-1)^2}-\frac{m_b^2 m_c^2}{x}\Bigg] Q^2\\
\nonumber&&+\Bigg[\frac{6 (x-1) (y-1) Q^2 x y^3 m_c^2 z^5}{(z-1)^2}-\frac{3 Q^2 (x-1) x y^3 m_b m_c z^5}{(z-1)^2}+\frac{3 (x-1) Q^2 x y^3 m_b m_c z^4}{(z-1)^2}+\frac{3 (x-1) Q^2 x y^3 m_b m_c z^3}{(y-1)^2}\\
\nonumber&&-\frac{3 Q^2 (x-1) x y^4 m_b m_c z^3}{(y-1)^2}-\frac{3 (y-1) y m_b m_c^3 z^3}{(z-1)^2}+\frac{6 (x-1) (z-1) Q^2 x y^3 m_b^2 z^2}{(y-1)^2}+\frac{6 (y-1) (z-1) Q^2 x m_c^2 z^2}{(x-1)^2}\\
\nonumber&&+\frac{3 (y-1) (z-1) Q^2 x m_b m_c z^2}{(x-1)^2}-\frac{3 Q^2 (y-1) (z-1) m_b m_c z^2}{(x-1)^2}-\frac{3 Q^2 (y-1) (z-1) m_b m_c z^2}{x}-\frac{4 y m_b^2 m_c^2 z^3}{(z-1)^2}\\
\nonumber&&-\frac{6 Q^2 (x-1) (y-1) (z-1) m_b^2 z^2}{x^2}-\frac{4 x m_b^2 m_c^2 z}{(x-1)^2}-\frac{4 y^2 m_b^2 m_c^2 z}{(y-1)^2}+\frac{3 x m_b m_c^3 z}{(x-1)^2}+\frac{4 y m_b^2 m_c^2 z}{(y-1)^2}+\frac{4 m_b^2 m_c^2 z}{(x-1)^2}+\frac{4 m_b^2 m_c^2 z}{x}\\
\nonumber&&+3 \left(\frac{3 (x-1) x z^2 m_b m_c y^4}{(y-1)^2}+\frac{2 (x-1) (z-1) x z m_b^2 y^3}{(y-1)^2}+\frac{2 (x-1) (y-1) x z^4 m_c^2 y^3}{(z-1)^2}\right.\\
\nonumber&&+\frac{3 (x-1) x z^4 m_b m_c y^3}{(z-1)^2}+\frac{2 (x-1) x z m_b m_c y^3}{(y-1)^2}-\frac{3 (x-1) x z^2 m_b m_c y^3}{(y-1)^2}-\frac{(x-1) x z^3 m_b m_c y^3}{(z-1)^2}\\
\nonumber&&+\frac{2 (y-1) (z-1) x z m_c^2}{(x-1)^2}+\frac{3 (y-1) (z-1) z m_b m_c}{(x-1)^2}+\frac{3 (y-1) (z-1) z m_b m_c}{x}-\frac{3 x (y-1) (z-1) z m_b m_c}{(x-1)^2}\\
\nonumber&&\left.-\frac{2 (x-1) (y-1) (z-1) z m_b^2}{x^2}-\frac{2 (y-1) (z-1) m_b m_c}{(x-1)^2 y}-\frac{2 (y-1) (z-1) m_b m_c}{x^2 y}\right) F\left(m_b,m_c,Q^2\right) z\\
\nonumber&&-\frac{3 (x-1) m_b^3 m_c z}{x^2}-\frac{4 y m_b^2 m_c^2}{(y-1)^2}-\frac{3 y (z-1) m_b^3 m_c}{(y-1)^2}-\frac{4 m_b^2 m_c^2}{(x-1)^2 y}-\frac{4 m_b^2 m_c^2}{x^2 y}\Bigg] \log \left(F\left(m_b,m_c,Q^2\right)\right)\Bigg\}\\
\nonumber&&+\frac{\langle g_{s}^{2}GG \rangle}{256 \pi ^6}\int_{0}^{1}dx\int_{0}^{1}dy\int_{0}^{1}dz\;\frac{3 c_2}{(x-1) x (y-1) y (z-1)} \log \left(F\left(m_b,m_c,Q^2\right)\right) \\
\nonumber&&\left\{-z F\left(m_b,m_c,Q^2\right)\left[m_b m_c \left(y^3 z \left(x^2 (6 z-2)+x (2-8 z)+3 z-1\right)+y^2 \left(x^2 \left(-6 z^2+6 z-2\right)+x \left(6 z^2-4 z+1\right)\right.\right.\right.\right.\\
\nonumber&&\left.\left.-3 z+2\right)+\left(2 x^4-4 x^3+3 x-1\right) y^4 z^2-4 y (z-1)^2+2 (z-1)^2\right)-\left(2 x^2-3 x+1\right) y^2 (z-1) m_b^2 ((y-2) z+1)\\
\nonumber&&\left.+(2 x-1) x (y-1) y^2 z m_c^2 ((y-2) z+1)-8 Q^2 (x-1) x (1-y) y^2 (z-1) z^2 \left((x-1) x y^2 z-y z+y+z-1\right)\right]\\
\nonumber&&-2 Q^2 z m_b m_c \left[(x-1)^2 x^2 y^4 z^2+y^2 (z-1)^2-2 y (z-1)^2+(z-1)^2\right]+2 m_b^2 m_c^2 \left[(x-1) x y^2 z-y z+y+z-1\right]\\
\nonumber&&\left.-2 Q^4 (x-1) x (1-y) y^2 (z-1) z^3 \left[(x-1) x y^2 z-y z+y+z-1\right]\right\}\\
\nonumber\Pi_1^{pert}(Q^{2})&=& \frac{1}{3072 \pi ^6}\int_{0}^{1}dx\int_{0}^{1}dy\int_{0}^{1}dz\;c_1 y z^2 F\left(m_b,m_c,Q^2\right){}^2 \log \left(F\left(m_b,m_c,Q^2\right)\right)\\
\nonumber&&\Bigg\{-3 (x-1) x (y-1) y^2 (z-1) z^3 F\left(m_b,m_c,Q^2\right)^2+F\left(m_b,m_c,Q^2\right) \left(-6 (x-1) x y^2 z^2 m_b m_c+6 (y-1) (z-1) z m_b m_c\right.\\
\nonumber&&\left.\left.+20 Q^2 (x-1) x (y-1) y^2 (z-1) z^3\right)+12 Q^2 (x-1) x y^2 z^2 m_b m_c-12 Q^2 (y-1) (z-1) z m_b m_c+12 m_b^2 m_c^2\right.\\
\nonumber&&-12 Q^4 (x-1) x (y-1) y^2 (z-1) z^3\Bigg\}
\\
\nonumber\Pi_1^{GG}(Q^{2})&=& \frac{\langle g_{s}^{2}GG \rangle}{18432 \pi ^6}\int_{0}^{1}dx\int_{0}^{1}dy\int_{0}^{1}dz\\
\nonumber&&c_1 \Bigg\{\frac{2}{F\left(m_b,m_c,Q^2\right)}\Bigg[\left(\frac{z-x z}{x^2}-\frac{y (z-1)}{(y-1)^2}\right) m_c m_b^3-\frac{1}{(x-1) x^2 (y-1)^2 (z-1)}\left(z \left((x-1)^2 (z-1)^2 \left(\left(x^3-1\right) y^3\right.\right.\right.\\
\nonumber&&\left.\left.\left.+3 y^2-3 y+1\right)z Q^2+(y-1) \left((x-1) x z y^2+\left(-x^2+x+z-1\right) y-z+1\right) x m_c^2\right) m_b^2\right)\\
\nonumber&&-\frac{1}{{(x-1)^2 x (y-1) (z-1)^2}}\left(z m_c \left((x-1) (z-1) \left((x-1)^2 x^2 z^2 y^4-(x-1)^2 x^2 z y^3-(z-1)^2 y^2+2 (z-1)^2 y\right.\right.\right.\\
\nonumber&&\left.-(z-1)^2\right) z Q^2\left.\left.+(y-1) \left((y-1) x^2 y z^2+(y-1) y z^2-x \left(\left(2 y^2-2 y+1\right) z^2-2 z+1\right)\right) x m_c^2\right) m_b\right)\\
\nonumber&&\left.-\frac{1}{{(x-1)^2 (z-1)^2}}\left(Q^2 x (y-1) z^2 \left(\left((x-1)^3 y^3+1\right) z^3-3 z^2+3 z-1\right) m_c^2\right)\Bigg] Q^2+\left[\frac{10 (x-1) (y-1) Q^2 x y^3 m_c^2 z^5}{(z-1)^2}\right.\right.\\
\nonumber&&+\frac{11 (x-1) Q^2 x y^3 m_b m_c z^5}{(z-1)^2}-\frac{7 Q^2 (x-1) x y^3 m_b m_c z^4}{(z-1)^2}+\frac{3 (y-1) y m_b m_c^3 z^3}{(z-1)^2}+\frac{4 y m_b^2 m_c^2 z^3}{(z-1)^2}+\frac{11 (x-1) Q^2 x y^4 m_b m_c z^3}{(y-1)^2}\\
\nonumber&&-\frac{11 Q^2 (x-1) x y^3 m_b m_c z^3}{(y-1)^2}+\frac{10 (x-1) (z-1) Q^2 x y^3 m_b^2 z^2}{(y-1)^2}+\frac{10 (y-1) (z-1) Q^2 x m_c^2 z^2}{(x-1)^2}+\frac{4 (x-1) Q^2 x y^3 m_b m_c z^2}{(y-1)^2}\\
\nonumber&&+\frac{11 (y-1) (z-1) Q^2 m_b m_c z^2}{(x-1)^2}+\frac{11 (y-1) (z-1) Q^2 m_b m_c z^2}{x}-\frac{11 Q^2 x (y-1) (z-1) m_b m_c z^2}{(x-1)^2}\\
\nonumber&&-\frac{10 Q^2 (x-1) (y-1) (z-1) m_b^2 z^2}{x^2}+\frac{4 y^2 m_b^2 m_c^2 z}{(y-1)^2}+\frac{4 x m_b^2 m_c^2 z}{(x-1)^2}+3 \left(-\frac{3 (x-1) x z^2 m_b m_c y^4}{(y-1)^2}+\frac{(x-1) x z^3 m_b m_c y^3}{(z-1)^2}\right.\\
\nonumber&&+\frac{3 (x-1) x z^2 m_b m_c y^3}{(y-1)^2}-\frac{2 (x-1) x (z-1) z m_b^2 y^3}{(y-1)^2}-\frac{2 (x-1) x z m_b m_c y^3}{(y-1)^2}-\frac{2 (x-1) x (y-1) z^4 m_c^2 y^3}{(z-1)^2}\\
\nonumber&&-\frac{3 (x-1) x z^4 m_b m_c y^3}{(z-1)^2}+\frac{2 (x-1) (y-1) (z-1) z m_b^2}{x^2}+\frac{3 (y-1) (z-1) x z m_b m_c}{(x-1)^2}+\frac{2 (y-1) (z-1) m_b m_c}{(x-1)^2 y}\\
\nonumber&&\left.+\frac{2 (y-1) (z-1) m_b m_c}{x^2 y}-\frac{2 x (y-1) (z-1) z m_c^2}{(x-1)^2}-\frac{3 (y-1) (z-1) z m_b m_c}{(x-1)^2}-\frac{3 (y-1) (z-1) z m_b m_c}{x}\right) F\left(m_b,m_c,Q^2\right) z\\
\nonumber&&-\frac{4 m_b^2 m_c^2 z}{(x-1)^2}-\frac{4 y m_b^2 m_c^2 z}{(y-1)^2}-\frac{4 m_b^2 m_c^2 z}{x}-\frac{4 Q^2 (y-1) (z-1) m_b m_c z}{(x-1)^2 y}-\frac{4 Q^2 (y-1) (z-1) m_b m_c z}{x^2 y}+\frac{4 y m_b^2 m_c^2}{(y-1)^2}+\frac{4 m_b^2 m_c^2}{(x-1)^2 y}\\
\nonumber&&\left.+\frac{4 m_b^2 m_c^2}{x^2 y}+\frac{3 (z-1) y m_b^3 m_c}{(y-1)^2}+\frac{3 (x-1) m_b^3 m_c z}{x^2}-\frac{3 x m_b m_c^3 z}{(x-1)^2}\right] \log \left(F\left(m_b,m_c,Q^2\right)\right)\Bigg\}\\
\nonumber&&+\frac{\langle g_{s}^{2}GG \rangle}{256 \pi ^6}\int_{0}^{1}dx\int_{0}^{1}dy\int_{0}^{1}dz\;\frac{c_2}{(x-1) x (y-1) y (z-1)} \log \left(F\left(m_b,m_c,Q^2\right)\right)\\
\nonumber&&\left\{z F\left(m_b,m_c,Q^2\right) \left(3 m_b m_c \left(y^3 z \left(x^2 (6 z-2)+x (2-8 z)+3 z-1\right)+y^2 \left(x^2 \left(-6 z^2+6 z-2\right)+x \left(6 z^2-4 z+1\right)\right.\right.\right.\right.\\
\nonumber&&\left.\left.-3 z+2\right)+\left(2 x^4-4 x^3+3 x-1\right) y^4 z^2-4 y (z-1)^2+2 (z-1)^2\right)-3 \left(2 x^2-3 x+1\right) y^2 (z-1) m_b^2 ((y-2) z+1)\\
\nonumber&&\left.+3 (2 x-1) x (y-1) y^2 z m_c^2 ((y-2) z+1)+8 Q^2 (x-1) x (1-y) y^2 (z-1) z^2 \left((x-1) x y^2 z-y z+y+z-1\right)\right)\\
\nonumber&&-2 Q^2 z m_b m_c \left(y^3 z \left(x^2 (6 z-2)+x (2-8 z)+3 z-1\right)+y^2 \left(x^2 \left(-6 z^2+6 z-2\right)+x \left(6 z^2-4 z+1\right)+z^2-5 z+3\right)\right.\\
\nonumber&&\left.+\left(3 x^4-6 x^3+x^2+3 x-1\right) y^4 z^2-6 y (z-1)^2+3 (z-1)^2\right)-6 m_b^2 m_c^2 \left((x-1) x y^2 z-y z+y+z-1\right)\\
\nonumber&&+2 Q^2 \left(2 x^2-3 x+1\right) y^2 (z-1) z m_b^2 ((y-2) z+1)-2 Q^2 x (2 x-1) (y-1) y^2 z^2 m_c^2 ((y-2) z+1)\\
\nonumber&&\left.-2 Q^4 (x-1) x (1-y) y^2 (z-1) z^3 \left((x-1) x y^2 z-y z+y+z-1\right)\right\}\\
\end{eqnarray}
}
The correlation function for current $J_4$ and $j_4$ is shown as:
{\allowdisplaybreaks
\begin{eqnarray}
\nonumber
\Pi_0^{pert}(Q^{2})&=& \frac{1}{1024 \pi ^6}\int_{0}^{1}dx\int_{0}^{1}dy\int_{0}^{1}dz\;c_1 y z^2 F\left(m_b,m_c,Q^2\right)^2 \log \left(F\left(m_b,m_c,Q^2\right)\right)\\
\nonumber&&\left\{(x-1) x (y-1) y^2 (z-1) z^3 F\left(m_b,m_c,Q^2\right){}^2+2 z F\left(m_b,m_c,Q^2\right) \left((x-1) x y^2 z m_b m_c-(y-1) (z-1) m_b m_c\right.\right.\\
\nonumber&&\left.\left.+2 Q^2 (x-1) x (y-1) y^2 (z-1) z^2\right)-4 m_b^2 m_c^2-4 Q^4 (x-1) x (y-1) y^2 (z-1) z^3\right\}\\
\nonumber\Pi_0^{GG}(Q^{2})&=& \frac{\langle g_{s}^{2}GG \rangle}{18432 \pi ^6}\int_{0}^{1}dx\int_{0}^{1}dy\int_{0}^{1}dz\\
\nonumber&&c_1 \left\{\frac{2 z}{{F\left(m_b,m_c,Q^2\right)}}\left[-\frac{Q^2 (x-1) x (y-1) y^3 m_c^2 z^4}{(z-1)^2}+\frac{(x-1) (y-1) (z-1) Q^2 m_b^2 z}{x^2}-\frac{Q^2 x (y-1) (z-1) m_c^2 z}{(x-1)^2}\right.\right.\\
\nonumber&&\left.+\frac{y m_b^2 m_c^2 z^2}{(z-1)^2}-\frac{Q^2 (x-1) x y^3 (z-1) m_b^2 z}{(y-1)^2}-\frac{y m_b^2 m_c^2 z}{(z-1)^2}+\frac{y^2 m_b^2 m_c^2}{(y-1)^2}+\frac{x m_b^2 m_c^2}{(x-1)^2}-\frac{m_b^2 m_c^2}{(x-1)^2}-\frac{y m_b^2 m_c^2}{(y-1)^2}-\frac{m_b^2 m_c^2}{x}\right] Q^2\\
\nonumber&&+\left[\frac{6 (x-1) (y-1) Q^2 x y^3 m_c^2 z^5}{(z-1)^2}-\frac{3 Q^2 (x-1) x y^3 m_b m_c z^5}{(z-1)^2}+\frac{3 (x-1) Q^2 x y^3 m_b m_c z^4}{(z-1)^2}+\frac{3 (x-1) Q^2 x y^3 m_b m_c z^3}{(y-1)^2}\right.\\
\nonumber&&-\frac{3 Q^2 (x-1) x y^4 m_b m_c z^3}{(y-1)^2}-\frac{3 (y-1) y m_b m_c^3 z^3}{(z-1)^2}-\frac{4 y m_b^2 m_c^2 z^3}{(z-1)^2}+\frac{6 (x-1) (z-1) Q^2 x y^3 m_b^2 z^2}{(y-1)^2}+\frac{6 (y-1) (z-1) Q^2 x m_c^2 z^2}{(x-1)^2}\\
\nonumber&&+\frac{3 (y-1) (z-1) Q^2 x m_b m_c z^2}{(x-1)^2}-\frac{3 Q^2 (y-1) (z-1) m_b m_c z^2}{(x-1)^2}-\frac{3 Q^2 (y-1) (z-1) m_b m_c z^2}{x}-\frac{4 x m_b^2 m_c^2 z}{(x-1)^2}-\frac{4 y^2 m_b^2 m_c^2 z}{(y-1)^2}\\
\nonumber&&-\frac{6 Q^2 (x-1) (y-1) (z-1) m_b^2 z^2}{x^2}+\frac{3 x m_b m_c^3 z}{(x-1)^2}+\frac{4 y m_b^2 m_c^2 z}{(y-1)^2}+\frac{4 m_b^2 m_c^2 z}{(x-1)^2}+\frac{4 m_b^2 m_c^2 z}{x}-\frac{3 (x-1) m_b^3 m_c z}{x^2}-\frac{4 y m_b^2 m_c^2}{(y-1)^2}\\
\nonumber&&+3 \left(\frac{3 (x-1) x z^2 m_b m_c y^4}{(y-1)^2}+\frac{2 (x-1) (z-1) x z m_b^2 y^3}{(y-1)^2}+\frac{2 (x-1) (y-1) x z^4 m_c^2 y^3}{(z-1)^2}\right.\\
\nonumber&&+\frac{3 (x-1) x z^4 m_b m_c y^3}{(z-1)^2}+\frac{2 (x-1) x z m_b m_c y^3}{(y-1)^2}-\frac{3 (x-1) x z^2 m_b m_c y^3}{(y-1)^2}-\frac{(x-1) x z^3 m_b m_c y^3}{(z-1)^2}+\frac{2 (y-1) (z-1) x z m_c^2}{(x-1)^2}\\
\nonumber&&+\frac{3 (y-1) (z-1) z m_b m_c}{(x-1)^2}+\frac{3 (y-1) (z-1) z m_b m_c}{x}-\frac{3 x (y-1) (z-1) z m_b m_c}{(x-1)^2}-\frac{2 (x-1) (y-1) (z-1) z m_b^2}{x^2}\\
\nonumber&&\left.-\frac{2 (y-1) (z-1) m_b m_c}{(x-1)^2 y}-\frac{2 (y-1) (z-1) m_b m_c}{x^2 y}\right) F\left(m_b,m_c,Q^2\right) z\\
\nonumber&&\left.\left.-\frac{3 y (z-1) m_b^3 m_c}{(y-1)^2}-\frac{4 m_b^2 m_c^2}{(x-1)^2 y}-\frac{4 m_b^2 m_c^2}{x^2 y}\right] \log \left(F\left(m_b,m_c,Q^2\right)\right)\right\}\\
\nonumber&&+\frac{\langle g_{s}^{2}GG \rangle}{256 \pi ^6}\int_{0}^{1}dx\int_{0}^{1}dy\int_{0}^{1}dz\;\frac{3 c_2}{ (x-1) x (y-1) y (z-1)} \log \left(F\left(m_b,m_c,Q^2\right)\right)\\
\nonumber&&\left\{-z F\left(m_b,m_c,Q^2\right) \left(m_b m_c \left(y^3 z \left(x^2 (6-10 z)+6 x (2 z-1)-3 z+1\right)+y^2 \left(2 x^2 \left(5 z^2-5 z+1\right)+x \left(-10 z^2+8 z-1\right)\right.\right.\right.\right.\\
\nonumber&&\left.\left.+4 z^2-5 z+2\right)+\left(2 x^4-4 x^3+4 x^2-3 x+1\right) y^4 z^2-4 y (z-1)^2+2 (z-1)^2\right)+\left(2 x^2-3 x+1\right) y^2 (z-1) m_b^2 \left((y-2) z\right.\\
\nonumber&&\left.\left.+1\right)-x (2 x-1) (y-1) y^2 z m_c^2 ((y-2) z+1)-8 Q^2 (x-1) x (1-y) y^2 (z-1) z^2 \left((x-1) x y^2 z-y z+y+z-1\right)\right)\\
\nonumber&&-2 Q^2 z m_b m_c \left((x-1)^2 x^2 y^4 z^2+y^2 (z-1)^2-2 y (z-1)^2+(z-1)^2\right)+2 m_b^2 m_c^2 \left((x-1) x y^2 z-y z+y+z-1\right)\\
\nonumber&&\left.-2 Q^4 (x-1) x (1-y) y^2 (z-1) z^3 \left((x-1) x y^2 z-y z+y+z-1\right)\right\}\\
\nonumber\Pi_1^{pert}(Q^{2})&=& \frac{1}{3072 \pi ^6}\int_{0}^{1}dx\int_{0}^{1}dy\int_{0}^{1}dz\;c_1 y z^2 F\left(m_b,m_c,Q^2\right)^2 \log \left(F\left(m_b,m_c,Q^2\right)\right)\\
\nonumber&&\Bigg\{-3 (x-1) x (y-1) y^2 (z-1) z^3 F\left(m_b,m_c,Q^2\right)^2+F\left(m_b,m_c,Q^2\right) \left(-6 (x-1) x y^2 z^2 m_b m_c+6 (y-1) (z-1) z m_b m_c\right.\\
\nonumber&&\left.+20 Q^2 (x-1) x (y-1) y^2 (z-1) z^3\right)+12 Q^2 (x-1) x y^2 z^2 m_b m_c\\
\nonumber&&-12 Q^2 (y-1) (z-1) z m_b m_c+12 m_b^2 m_c^2-12 Q^4 (x-1) x (y-1) y^2 (z-1) z^3\Bigg\}\\
\nonumber\Pi_1^{GG}(Q^{2})&=& \frac{\langle g_{s}^{2}GG \rangle}{18432 \pi ^6}\int_{0}^{1}dx\int_{0}^{1}dy\int_{0}^{1}dz\\
\nonumber&&c_1 \left\{\frac{2}{F\left(m_b,m_c,Q^2\right)} \left[\left(\frac{z-x z}{x^2}-\frac{y (z-1)}{(y-1)^2}\right) m_c m_b^3-\frac{1}{(x-1) x^2 (y-1)^2 (z-1)}\left(z \left((x-1)^2 (z-1)^2 \left(\left(x^3-1\right) y^3+3 y^2\right.\right.\right.\right.\right.\\
\nonumber&&\left.\left.\left.\left.\left.-3 y+1\right)z Q^2+(y-1) \left((x-1) x z y^2+\left(-x^2+x+z-1\right) y-z+1\right) x m_c^2\right) m_b^2\right)-\frac{1}{(x-1)^2 x (y-1) (z-1)^2}\left( \left((x-1) \right.\right.\right.\right. \\
\nonumber&&(z-1)\left((x-1)^2 x^2 z^2 y^4-(x-1)^2 x^2 z y^3-(z-1)^2 y^2+2 (z-1)^2 y-(z-1)^2\right) z Q^2+(y-1) \left((y-1) x^2 y z^2+(y-1) y z^2\right.\\
\nonumber&&\left.\left.\left.\left.\left.-x \left(\left(2 y^2-2 y+1\right) z^2-2 z+1\right)\right) x m_c^2\right)z m_c m_b\right)-\frac{Q^2 x (y-1) z^2 \left(\left((x-1)^3 y^3+1\right) z^3-3 z^2+3 z-1\right) m_c^2}{(x-1)^2 (z-1)^2}\right] Q^2\right.\\
\nonumber&&+\left[\frac{10 (x-1) (y-1) Q^2 x y^3 m_c^2 z^5}{(z-1)^2}+\frac{11 (x-1) Q^2 x y^3 m_b m_c z^5}{(z-1)^2}-\frac{7 Q^2 (x-1) x y^3 m_b m_c z^4}{(z-1)^2}+\frac{3 (y-1) y m_b m_c^3 z^3}{(z-1)^2}\right.\\
\nonumber&&+\frac{4 y m_b^2 m_c^2 z^3}{(z-1)^2}-\frac{11 Q^2 x (y-1) (z-1) m_b m_c z^2}{(x-1)^2}+\frac{3 (x-1) m_b^3 m_c z}{x^2}-\frac{3 x m_b m_c^3 z}{(x-1)^2}+\frac{4 y^2 m_b^2 m_c^2 z}{(y-1)^2}+\frac{4 x m_b^2 m_c^2 z}{(x-1)^2}\\
\nonumber&&+\frac{11 (x-1) Q^2 x y^4 m_b m_c z^3}{(y-1)^2}-\frac{11 Q^2 (x-1) x y^3 m_b m_c z^3}{(y-1)^2}+\frac{10 (x-1) (z-1) Q^2 x y^3 m_b^2 z^2}{(y-1)^2}+\frac{10 (y-1) (z-1) Q^2 x m_c^2 z^2}{(x-1)^2}\\
\nonumber&&+\frac{4 (x-1) Q^2 x y^3 m_b m_c z^2}{(y-1)^2}+\frac{11 (y-1) (z-1) Q^2 m_b m_c z^2}{(x-1)^2}+\frac{11 (y-1) (z-1) Q^2 m_b m_c z^2}{x}+\frac{4 y m_b^2 m_c^2}{(y-1)^2}\\
\nonumber&&-\frac{10 Q^2 (x-1) (y-1) (z-1) m_b^2 z^2}{x^2}+3 \left(-\frac{3 (x-1) x z^2 m_b m_c y^4}{(y-1)^2}+\frac{(x-1) x z^3 m_b m_c y^3}{(z-1)^2}-\frac{3 (x-1) x z^4 m_b m_c y^3}{(z-1)^2}\right.\\
\nonumber&&+\frac{3 (x-1) x z^2 m_b m_c y^3}{(y-1)^2}-\frac{2 (x-1) x (z-1) z m_b^2 y^3}{(y-1)^2}-\frac{2 (x-1) x z m_b m_c y^3}{(y-1)^2}-\frac{2 (x-1) x (y-1) z^4 m_c^2 y^3}{(z-1)^2}\\
\nonumber&&+\frac{2 (x-1) (y-1) (z-1) z m_b^2}{x^2}+\frac{3 (y-1) (z-1) x z m_b m_c}{(x-1)^2}+\frac{2 (y-1) (z-1) m_b m_c}{(x-1)^2 y}+\frac{2 (y-1) (z-1) m_b m_c}{x^2 y}\\
\nonumber&&\left.-\frac{2 x (y-1) (z-1) z m_c^2}{(x-1)^2}-\frac{3 (y-1) (z-1) z m_b m_c}{(x-1)^2}-\frac{3 (y-1) (z-1) z m_b m_c}{x}\right) F\left(m_b,m_c,Q^2\right) z\\
\nonumber&&-\frac{4 m_b^2 m_c^2 z}{(x-1)^2}-\frac{4 y m_b^2 m_c^2 z}{(y-1)^2}-\frac{4 m_b^2 m_c^2 z}{x}-\frac{4 Q^2 (y-1) (z-1) m_b m_c z}{(x-1)^2 y}-\frac{4 Q^2 (y-1) (z-1) m_b m_c z}{x^2 y}+\frac{4 m_b^2 m_c^2}{(x-1)^2 y}\\
\nonumber&&\left.\left.+\frac{4 m_b^2 m_c^2}{x^2 y}+\frac{3 (z-1) y m_b^3 m_c}{(y-1)^2}\right] \log \left(F\left(m_b,m_c,Q^2\right)\right)\right\}\\
\nonumber&&+\frac{\langle g_{s}^{2}GG \rangle}{256 \pi ^6}\int_{0}^{1}dx\int_{0}^{1}dy\int_{0}^{1}dz\;\frac{c_2}{(x-1) x (y-1) y (z-1)} \log \left(F\left(m_b,m_c,Q^2\right)\right)\\
\nonumber&&\left\{z F\left(m_b,m_c,Q^2\right) \left(3 m_b m_c \left(y^3 z \left(x^2 (6-10 z)+6 x (2 z-1)-3 z+1\right)+y^2 \left(2 x^2 \left(5 z^2-5 z+1\right)\right.\right.\right.\right.\\
\nonumber&&\left.\left.+x \left(-10 z^2+8 z-1\right)+4 z^2-5 z+2\right)+\left(2 x^4-4 x^3+4 x^2-3 x+1\right) y^4 z^2-4 y (z-1)^2+2 (z-1)^2\right)\\
\nonumber&&+3 \left(2 x^2-3 x+1\right) y^2 (z-1) m_b^2 ((y-2) z+1)\\
\nonumber&&\left.-3 x (2 x-1) (y-1) y^2 z m_c^2 ((y-2) z+1)-8 Q^2 (x-1) x (y-1) y^2 (z-1) z^2 \left((x-1) x y^2 z-y z+y+z-1\right)\right)\\
\nonumber&&\left.-2 Q^2 z m_b m_c \left(y^3 z \left(x^2 (6-10 z)+6 x (2 z-1)-3 z+1\right)+y^2 \left(2 x^2 \left(5 z^2-5 z+1\right)+x \left(-10 z^2+8 z-1\right)+5 z^2-7 z+3\right)\right.\right.\\
\nonumber&&\left.+\left(3 x^4-6 x^3+5 x^2-3 x+1\right) y^4 z^2-6 y (z-1)^2+3 (z-1)^2\right)-6 m_b^2 m_c^2 \left((x-1) x y^2 z-y z+y+z-1\right)\\
\nonumber&&-2 Q^2 \left(2 x^2-3 x+1\right) y^2 (z-1) z m_b^2 ((y-2) z+1)+2 Q^2 (2 x-1) x (y-1) y^2 z^2 m_c^2 ((y-2) z+1)\\
\nonumber&&\left.+2 Q^4 (x-1) x (y-1) y^2 (z-1) z^3 \left((x-1) x y^2 z-y z+y+z-1\right)\right\}\\
\end{eqnarray}
}
The correlation function for current $J_5$ and $j_5$ is shown as:
{\allowdisplaybreaks
\begin{eqnarray}
\nonumber
\Pi_{0}^{pert}(Q^{2})&=& \frac{1}{1024 \pi ^6}\int_{0}^{1}dx\int_{0}^{1}dy\int_{0}^{1}dz\;c_1 y z^2 F\left(m_b,m_c,Q^2\right){}^2 \log \left(F\left(m_b,m_c,Q^2\right)\right)\\
\nonumber&&\left\{(x-1) x (y-1) y^2 (z-1) z^3 F\left(m_b,m_c,Q^2\right)^2+2 z F\left(m_b,m_c,Q^2\right) \left((x-1) x y^2 z m_b m_c-(y-1) (z-1) m_b m_c\right.\right.\\
\nonumber&&\left.+2 Q^2 (x-1) x (y-1) y^2 (z-1) z^2\right)+3 Q^2 (x-1) x y^2 z^2 m_b m_c-3 Q^2 (y-1) (z-1) z m_b m_c-6 m_b^2 m_c^2\\
\nonumber&&\left.+6 Q^4 (x-1) x (y-1) y^2 (z-1) z^3\right\}\\
\nonumber\Pi_{0T}^{pert}(Q^{2})&=& \frac{1}{3072 \pi ^6}\int_{0}^{1}dx\int_{0}^{1}dy\int_{0}^{1}dz\;c_1 y z^2 F\left(m_b,m_c,Q^2\right){}^2 \log \left(F\left(m_b,m_c,Q^2\right)\right)\\
\nonumber&&\left\{3 (x-1) x (y-1) y^2 (z-1) z^3 F\left(m_b,m_c,Q^2\right){}^2+2 z F\left(m_b,m_c,Q^2\right) \left((x-1) x y^2 z m_b m_c-(y-1) (z-1) m_b m_c\right.\right.\\
\nonumber&&\left.-6 Q^2 (x-1) x (y-1) y^2 (z-1) z^2\right)-3 Q^2 (x-1) x y^2 z^2 m_b m_c+3 Q^2 (y-1) (z-1) z m_b m_c-6 m_b^2 m_c^2\\
\nonumber&&\left.+6 Q^4 (x-1) x (y-1) y^2 (z-1) z^3\right\}\\
\nonumber\Pi_{1}^{pert}(Q^{2})&=& \frac{1}{128 \pi ^6}\int_{0}^{1}dx\int_{0}^{1}dy\int_{0}^{1}dz\;c_1 y z^2 F\left(m_b,m_c,Q^2\right){}^2 \log \left(F\left(m_b,m_c,Q^2\right)\right)\\
\nonumber&&\left\{-(x-1) x (y-1) y^2 (z-1) z^3 F\left(m_b,m_c,Q^2\right){}^2-2 (x-1) x y^2 z^2 m_b m_c F\left(m_b,m_c,Q^2\right)\right.\\
\nonumber&&\left.+2 (y-1) (z-1) z m_b m_c F\left(m_b,m_c,Q^2\right)+6 m_b^2 m_c^2+6 Q^4 (x-1) x (y-1) y^2 (z-1) z^3\right\}\\
\nonumber\Pi_{2}^{pert}(Q^{2})&=& \frac{1}{768 \pi ^6}\int_{0}^{1}dx\int_{0}^{1}dy\int_{0}^{1}dz\;5 c_1 y z^2 F\left(m_b,m_c,Q^2\right){}^2 \log \left(F\left(m_b,m_c,Q^2\right)\right)\\
\nonumber&& \left\{(x-1) x (y-1) y^2 (z-1) z^3 F\left(m_b,m_c,Q^2\right){}^2+2 z F\left(m_b,m_c,Q^2\right) \left((x-1) x y^2 z m_b m_c-(y-1) (z-1) m_b m_c\right.\right.\\
\nonumber&&\left.-4 Q^2 (x-1) x (y-1) y^2 (z-1) z^2\right)-6 Q^2 (x-1) x y^2 z^2 m_b m_c+6 Q^2 (y-1) (z-1) z m_b m_c-6 m_b^2 m_c^2\\
\nonumber&&\left.+6 Q^4 (x-1) x (y-1) y^2 (z-1) z^3\right\}\\
\nonumber\Pi_0^{GG}(Q^{2})&=& \frac{\langle g_{s}^{2}GG \rangle}{12288 \pi ^6}\int_{0}^{1}dx\int_{0}^{1}dy\int_{0}^{1}dz\\
\nonumber&& c_1 \left\{\frac{1}{F\left(m_b,m_c,Q^2\right)}\Bigg[\left(\frac{z-x z}{x^2}-\frac{y (z-1)}{(y-1)^2}\right) m_c m_b^3+\frac{1}{(x-1) x^2 (y-1)^2 (z-1)}\left(2 z \left((x-1)^2 (z-1)^2 \left(\left(x^3-1\right) y^3\right.\right.\right.\right.\\
\nonumber&&\left.\left.\left.\left.\left.+3 y^2-3 y+1\right) z Q^2+(y-1) \left((x-1) x z y^2+\left(-x^2+x+z-1\right) y-z+1\right) x m_c^2\right) m_b^2\right)\right.\right.\\
\nonumber&&\left.\left.+\frac{2 (y-1)}{(x-1)^2 (z-1)^2}\left(\left((x-1)^3 y^3+1\right) z^3-3 z^2+3 z-1\right)\right.\right.\\
\nonumber&&\left.\left.Q^2 x z^2 m_c^2-\frac{z m_c}{(x-1)^2 x (y-1) (z-1)^2}\left((x-1) (z-1) \left((x-1)^2 x^2 z^2 y^4-(x-1)^2 x^2 z y^3-(z-1)^2 y^2+2 (z-1)^2 y\right.\right.\right.\right.\\
\nonumber&&\left.\left.\left.-(z-1)^2\right) z Q^2+(y-1) \left((y-1) x^2 y z^2+(y-1) y z^2-x \left(\left(2 y^2-2 y+1\right) z^2-2 z+1\right)\right) x m_c^2\right) m_b\Bigg] Q^2\right.\\
\nonumber&&\left.+2 \left[\frac{2 (x-1) (y-1) Q^2 x y^3 m_c^2 z^5}{(z-1)^2}+\frac{(x-1) Q^2 x y^3 m_b m_c z^5}{(z-1)^2}+\frac{(x-1) Q^2 x y^4 m_b m_c z^3}{(y-1)^2}-\frac{Q^2 (x-1) x y^3 m_b m_c z^3}{(y-1)^2}\right.\right.\\
\nonumber&&-\frac{(y-1) y m_b m_c^3 z^3}{(z-1)^2}-\frac{2 y m_b^2 m_c^2 z^3}{(z-1)^2}+\frac{2 (x-1) (z-1) Q^2 x y^3 m_b^2 z^2}{(y-1)^2}+\frac{2 (y-1) (z-1) Q^2 x m_c^2 z^2}{(x-1)^2}+\frac{(x-1) Q^2 x y^3 m_b m_c z^2}{(y-1)^2}\\
\nonumber&&+\frac{(y-1) (z-1) Q^2 m_b m_c z^2}{(x-1)^2}+\frac{(y-1) (z-1) Q^2 m_b m_c z^2}{x}-\frac{Q^2 x (y-1) (z-1) m_b m_c z^2}{(x-1)^2}\\
\nonumber&&+\frac{x m_b m_c^3 z}{(x-1)^2}+\frac{2 y m_b^2 m_c^2 z}{(y-1)^2}+\frac{2 m_b^2 m_c^2 z}{(x-1)^2}+\frac{2 m_b^2 m_c^2 z}{x}+\left(\frac{3 (x-1) x z^2 m_b m_c y^4}{(y-1)^2}+\frac{2 (x-1) (z-1) x z m_b^2 y^3}{(y-1)^2}\right.\\
\nonumber&&+\frac{3 (x-1) x z^4 m_b m_c y^3}{(z-1)^2}+\frac{2 (x-1) x z m_b m_c y^3}{(y-1)^2}-\frac{3 (x-1) x z^2 m_b m_c y^3}{(y-1)^2}-\frac{(x-1) x z^3 m_b m_c y^3}{(z-1)^2}+\frac{2 (y-1) (z-1) x z m_c^2}{(x-1)^2}\\
\nonumber&&+\frac{3 (y-1) (z-1) z m_b m_c}{(x-1)^2}+\frac{3 (y-1) (z-1) z m_b m_c}{x}-\frac{3 x (y-1) (z-1) z m_b m_c}{(x-1)^2}-\frac{2 (x-1) (y-1) (z-1) z m_b^2}{x^2}\\
\nonumber&&\left.+\frac{2 (x-1) (y-1) x z^4 m_c^2 y^3}{(z-1)^2}-\frac{2 (y-1) (z-1) m_b m_c}{(x-1)^2 y}-\frac{2 (y-1) (z-1) m_b m_c}{x^2 y}\right) F\left(m_b,m_c,Q^2\right) z-\frac{(x-1) m_b^3 m_c z}{x^2}\\
\nonumber&&-\frac{Q^2 (y-1) (z-1) m_b m_c z}{(x-1)^2 y}-\frac{Q^2 (y-1) (z-1) m_b m_c z}{x^2 y}-\frac{2 y m_b^2 m_c^2}{(y-1)^2}-\frac{y (z-1) m_b^3 m_c}{(y-1)^2}\\
\nonumber&&\left.\left.-\frac{2 m_b^2 m_c^2}{(x-1)^2 y}-\frac{2 m_b^2 m_c^2}{x^2 y}-\frac{2 x m_b^2 m_c^2 z}{(x-1)^2}-\frac{2 y^2 m_b^2 m_c^2 z}{(y-1)^2}\right] \log \left(F\left(m_b,m_c,Q^2\right)\right)\right\}\\
\nonumber &&-\frac{2 Q^2 (x-1) (y-1) (z-1) m_b^2 z^2}{x^2}+\frac{\langle g_{s}^{2}GG \rangle}{512 \pi ^6}\int_{0}^{1}dx\int_{0}^{1}dy\int_{0}^{1}dz\;3 c_2 \log \left(F\left(m_b,m_c,Q^2\right)\right) \\
\nonumber &&\left\{\left(4 x^2-4 x-3\right) y^3 z^4 F\left(m_b,m_c,Q^2\right){}^2+Q^2 y z^3 \left(\left(-4 x^2+4 x+1\right) y^2 z+y (4 z-5)-4 z+4\right) F\left(m_b,m_c,Q^2\right)\right.\\
\nonumber &&\left.+Q^2 y z^3 \left(\left(4 x^2-4 x-3\right) y^2 z+y (7-4 z)+4 (z-1)\right) F\left(m_b,m_c,Q^2\right)-\frac{1}{(x-1) x (y-1) (z-1)}\left(2 z m_b m_c \left(y^2 z\right.\right.\right.\\
\nonumber &&\left.\left.\left.\left(x^2 (6 z-7)\right.+x (7-6 z)-z+1\right)-y (z-1) \left(\left(6 x^2-6 x-1\right) z-1\right)+(x-1) x y^3 z^2-z+1\right) F\left(m_b,m_c,Q^2\right)\right)\\
\nonumber &&+y^2 z^3 (7-4 z) F\left(m_b,m_c,Q^2\right){}^2+4 y (z-1) z^3 F\left(m_b,m_c,Q^2\right){}^2+\frac{1}{(x-1) x (y-1) (z-1)}Q^2 z m_b m_c \left(y^2 z \left(x^2 (10 z-9)\right.\right.\\
\nonumber &&\left.\left.\left.+x (9-10 z)+z-1\right)-y (z-1) \left(10 x^2 z-10 x z+z+1\right)-(x-1) x y^3 z^2+z-1\right)+\frac{2 m_b^2 m_c^2 (y z-1)}{(x-1) x (y-1) (z-1)}\right.\\
\nonumber &&\left.+Q^4 y z^3 \left(\left(-4 x^2+4 x+1\right) y^2 z+y (4 z-5)-4 z+4\right)\right\}\\
\nonumber\Pi_{0T}^{GG}(Q^{2})&=& \frac{\langle g_{s}^{2}GG \rangle}{36864 \pi ^6}\int_{0}^{1}dx\int_{0}^{1}dy\int_{0}^{1}dz\\
\nonumber&&c_1 \left\{\frac{1}{F\left(m_b,m_c,Q^2\right)}\left[\frac{2 (x-1) (y-1) Q^2 x y^3 m_c^2 z^5}{(z-1)^2}+\frac{(x-1) Q^2 x y^3 m_b m_c z^5}{(z-1)^2}-\frac{Q^2 (x-1) x y^3 m_b m_c z^4}{(z-1)^2}+\frac{(y-1) y m_b m_c^3 z^3}{(z-1)^2}\right.\right.\\
\nonumber&&+\frac{(x-1) Q^2 x y^4 m_b m_c z^3}{(y-1)^2}-\frac{Q^2 (x-1) x y^3 m_b m_c z^3}{(y-1)^2}+\frac{2 (x-1) (z-1) Q^2 x y^3 m_b^2 z^2}{(y-1)^2}+\frac{2 (y-1) (z-1) Q^2 x m_c^2 z^2}{(x-1)^2}\\
\nonumber&&+\frac{2 y m_b^2 m_c^2 z^3}{(z-1)^2}+\frac{(y-1) (z-1) Q^2 m_b m_c z^2}{(x-1)^2}+\frac{(y-1) (z-1) Q^2 m_b m_c z^2}{x}-\frac{Q^2 x (y-1) (z-1) m_b m_c z^2}{(x-1)^2}-\frac{2 y m_b^2 m_c^2 z^2}{(z-1)^2}\\
\nonumber&&-\frac{2 Q^2 (x-1) (y-1) (z-1) m_b^2 z^2}{x^2}+\frac{2 y^2 m_b^2 m_c^2 z}{(y-1)^2}+\frac{2 x m_b^2 m_c^2 z}{(x-1)^2}+\frac{(x-1) m_b^3 m_c z}{x^2}-\frac{x m_b m_c^3 z}{(x-1)^2}-\frac{2 m_b^2 m_c^2 z}{(x-1)^2}-\frac{2 y m_b^2 m_c^2 z}{(y-1)^2}\\
\nonumber&&\left.\left.-\frac{2 m_b^2 m_c^2 z}{x}+\frac{(z-1) y m_b^3 m_c}{(y-1)^2}\right] Q^2+2 \left[-\frac{6 Q^2 (x-1) x (y-1) y^3 m_c^2 z^5}{(z-1)^2}-\frac{3 Q^2 (x-1) x y^3 m_b m_c z^5}{(z-1)^2}+\frac{2 m_b^2 m_c^2 z}{(x-1)^2}+\frac{2 m_b^2 m_c^2 z}{x}\right.\right.\\
\nonumber&&+\frac{3 (x-1) Q^2 x y^3 m_b m_c z^3}{(y-1)^2}-\frac{3 Q^2 (x-1) x y^4 m_b m_c z^3}{(y-1)^2}-\frac{(y-1) y m_b m_c^3 z^3}{(z-1)^2}-\frac{2 y m_b^2 m_c^2 z^3}{(z-1)^2}+\frac{6 (x-1) (y-1) (z-1) Q^2 m_b^2 z^2}{x^2}\\
\nonumber&&+\frac{3 (y-1) (z-1) Q^2 x m_b m_c z^2}{(x-1)^2}-\frac{6 Q^2 x (y-1) (z-1) m_c^2 z^2}{(x-1)^2}-\frac{3 Q^2 (y-1) (z-1) m_b m_c z^2}{(x-1)^2}-\frac{6 Q^2 (x-1) x y^3 (z-1) m_b^2 z^2}{(y-1)^2}\\
\nonumber&&-\frac{Q^2 (x-1) x y^3 m_b m_c z^2}{(y-1)^2}-\frac{3 Q^2 (y-1) (z-1) m_b m_c z^2}{x}+\frac{x m_b m_c^3 z}{(x-1)^2}+\frac{2 y m_b^2 m_c^2 z}{(y-1)^2}+\frac{2 (x-1) Q^2 x y^3 m_b m_c z^4}{(z-1)^2}\\
\nonumber&&+\left(\frac{3 (x-1) x z^2 m_b m_c y^4}{(y-1)^2}+\frac{6 (x-1) (z-1) x z m_b^2 y^3}{(y-1)^2}+\frac{6 (x-1) (y-1) x z^4 m_c^2 y^3}{(z-1)^2}+\frac{3 (x-1) x z^4 m_b m_c y^3}{(z-1)^2}\right.\\
\nonumber&&+\frac{2 (x-1) x z m_b m_c y^3}{(y-1)^2}-\frac{3 (x-1) x z^2 m_b m_c y^3}{(y-1)^2}-\frac{(x-1) x z^3 m_b m_c y^3}{(z-1)^2}-\frac{2 (y-1) (z-1) m_b m_c}{(x-1)^2 y}\\
\nonumber&&+\frac{6 (y-1) (z-1) x z m_c^2}{(x-1)^2}+\frac{3 (y-1) (z-1) z m_b m_c}{(x-1)^2}+\frac{3 (y-1) (z-1) z m_b m_c}{x}-\frac{2 (y-1) (z-1) m_b m_c}{x^2 y}\\
\nonumber&&\left.-\frac{3 x (y-1) (z-1) z m_b m_c}{(x-1)^2}-\frac{6 (x-1) (y-1) (z-1) z m_b^2}{x^2}\right) F\left(m_b,m_c,Q^2\right) z\\
\nonumber&&+\frac{(y-1) (z-1) Q^2 m_b m_c z}{(x-1)^2 y}+\frac{(y-1) (z-1) Q^2 m_b m_c z}{x^2 y}-\frac{2 x m_b^2 m_c^2 z}{(x-1)^2}-\frac{2 y^2 m_b^2 m_c^2 z}{(y-1)^2}-\frac{(x-1) m_b^3 m_c z}{x^2}-\frac{2 y m_b^2 m_c^2}{(y-1)^2}\\
\nonumber&&\left.\left.-\frac{y (z-1) m_b^3 m_c}{(y-1)^2}-\frac{2 m_b^2 m_c^2}{(x-1)^2 y}-\frac{2 m_b^2 m_c^2}{x^2 y}\right] \log \left(F\left(m_b,m_c,Q^2\right)\right)\right\}\\
\nonumber&&+\frac{\langle g_{s}^{2}GG \rangle}{512 \pi ^6}\int_{0}^{1}dx\int_{0}^{1}dy\int_{0}^{1}dz\;-\frac{3 c_2}{(x-1) x (y-1) (z-1)} \log \left(F\left(m_b,m_c,Q^2\right)\right)\\
\nonumber&& \left\{F\left(m_b,m_c,Q^2\right) \left(6 Q^2 (x-1) x (y-1) y^2 (z-1) z^3 (y z-1)-2 z m_b m_c \left(-y^2 z \left(x^2 (2 z-1)-2 x z+x+z-1\right)\right.\right.\right.\\
\nonumber&&\left.\left.+y (z-1) \left(2 x^2 z-2 x z+z+1\right)+(x-1) x y^3 z^2-z+1\right)\right)-3 (x-1) x (y-1) y^2 (z-1) z^3 (y z-1) F\left(m_b,m_c,Q^2\right){}^2\\
\nonumber&&+Q^2 z m_b m_c \left(-y^2 z \left(x^2 (2 z-1)-2 x z+x+z-1\right)+y (z-1) \left(2 x^2 z-2 x z+z+1\right)+(x-1) x y^3 z^2-z+1\right)\\
\nonumber&&\left.+2 m_b^2 m_c^2 (y z-1)+Q^4 (-(x-1)) x (y-1) y^2 (z-1) z^3 (y z-1)\right\}\\
\nonumber\Pi_1^{GG}(Q^{2})&=& \frac{\langle g_{s}^{2}GG \rangle}{768 \pi ^6}\int_{0}^{1}dx\int_{0}^{1}dy\int_{0}^{1}dz\\
\nonumber&&c_1 \left\{\frac{z}{F\left(m_b,m_c,Q^2\right)} \left[\frac{(x-1) (y-1) Q^2 x y^3 m_c^2 z^4}{(z-1)^2}-\frac{y m_b^2 m_c^2 z^2}{(z-1)^2}+\frac{(x-1) (z-1) Q^2 x y^3 m_b^2 z}{(y-1)^2}+\frac{y m_b^2 m_c^2 z}{(z-1)^2}\right.\right.\\
\nonumber&&\left.+\frac{(y-1) (z-1) Q^2 x m_c^2 z}{(x-1)^2}-\frac{Q^2 (x-1) (y-1) (z-1) m_b^2 z}{x^2}+\frac{y m_b^2 m_c^2}{(y-1)^2}+\frac{m_b^2 m_c^2}{(x-1)^2}+\frac{m_b^2 m_c^2}{x}-\frac{x m_b^2 m_c^2}{(x-1)^2}-\frac{y^2 m_b^2 m_c^2}{(y-1)^2}\right] Q^2\\
\nonumber&&+\left[\frac{(x-1) Q^2 x y^3 m_b m_c z^5}{(z-1)^2}-\frac{Q^2 (x-1) x y^3 m_b m_c z^4}{(z-1)^2}+\frac{(y-1) y m_b m_c^3 z^3}{(z-1)^2}+\frac{2 y m_b^2 m_c^2 z^3}{(z-1)^2}+\frac{(x-1) Q^2 x y^4 m_b m_c z^3}{(y-1)^2}\right.\\
\nonumber&&-\frac{Q^2 (x-1) x y^3 m_b m_c z^3}{(y-1)^2}+\frac{(y-1) (z-1) Q^2 m_b m_c z^2}{(x-1)^2}+\frac{(y-1) (z-1) Q^2 m_b m_c z^2}{x}-\frac{Q^2 x (y-1) (z-1) m_b m_c z^2}{(x-1)^2}\\
\nonumber&&+\frac{2 y^2 m_b^2 m_c^2 z}{(y-1)^2}+\frac{2 x m_b^2 m_c^2 z}{(x-1)^2}-\frac{x m_b m_c^3 z}{(x-1)^2}+\left(-\frac{3 (x-1) x z^2 m_b m_c y^4}{(y-1)^2}+\frac{(x-1) x z^3 m_b m_c y^3}{(z-1)^2}+\frac{3 (x-1) x z^2 m_b m_c y^3}{(y-1)^2}\right.\\
\nonumber&&-\frac{2 (x-1) x (z-1) z m_b^2 y^3}{(y-1)^2}-\frac{2 (x-1) x z m_b m_c y^3}{(y-1)^2}-\frac{2 (x-1) x (y-1) z^4 m_c^2 y^3}{(z-1)^2}-\frac{3 (x-1) x z^4 m_b m_c y^3}{(z-1)^2}\\
\nonumber&&+\frac{2 (x-1) (y-1) (z-1) z m_b^2}{x^2}+\frac{3 (y-1) (z-1) x z m_b m_c}{(x-1)^2}+\frac{2 (y-1) (z-1) m_b m_c}{(x-1)^2 y}+\frac{2 (y-1) (z-1) m_b m_c}{x^2 y}\\
\nonumber&&\left.-\frac{2 x (y-1) (z-1) z m_c^2}{(x-1)^2}-\frac{3 (y-1) (z-1) z m_b m_c}{(x-1)^2}-\frac{3 (y-1) (z-1) z m_b m_c}{x}\right) F\left(m_b,m_c,Q^2\right) z+\frac{(x-1) m_b^3 m_c z}{x^2}\\
\nonumber&&\left.\left.-\frac{2 m_b^2 m_c^2 z}{(x-1)^2}-\frac{2 y m_b^2 m_c^2 z}{(y-1)^2}-\frac{2 m_b^2 m_c^2 z}{x}+\frac{2 y m_b^2 m_c^2}{(y-1)^2}+\frac{2 m_b^2 m_c^2}{(x-1)^2 y}+\frac{2 m_b^2 m_c^2}{x^2 y}+\frac{(z-1) y m_b^3 m_c}{(y-1)^2}\right] \log \left(F\left(m_b,m_c,Q^2\right)\right)\right\}\\
\nonumber&&+\frac{\langle g_{s}^{2}GG \rangle}{64 \pi ^6}\int_{0}^{1}dx\int_{0}^{1}dy\int_{0}^{1}dz\;3 c_2 \log \left(F\left(m_b,m_c,Q^2\right)\right) \\
\nonumber&&\Bigg\{y z^3 \left(\left(-4 x^2+4 x+3\right) y^2 z+y (4 z-7)-4 z+4\right) F\left(m_b,m_c,Q^2\right){}^2\\
\nonumber&&\left.+F\left(m_b,m_c,Q^2\right) \left(2 z m_b m_c \left[\frac{(6 x-7) y z}{x-1}+\frac{y z-1}{x}+\frac{1}{x-1}+y^2 z \left(\frac{1}{y-1}+\frac{z}{z-1}\right)\right]\right.\right.\\
\nonumber&&\left.\left.+2 Q^2 y z^3 \left(\left(2 x^2-2 x-1\right) y^2 z+y (3-2 z)\right.\right.\right.\left.\left.+2 (z-1)\right)\right)-8 Q^2 y z^2 m_b m_c\\
\nonumber&&-\frac{2 m_b^2 m_c^2 (y z-1)}{(x-1) x (y-1) (z-1)}-2 Q^4 y z^3 \left((x-1) x y^2 z-y z+y+z-1\right)\Bigg\}\\
\nonumber\Pi_2^{GG}(Q^{2})&=& \frac{\langle g_{s}^{2}GG \rangle}{4608 \pi ^6}\int_{0}^{1}dx\int_{0}^{1}dy\int_{0}^{1}dz\\
\nonumber&&5 c_1 \left\{\frac{1}{F\left(m_b,m_c,Q^2\right)}\left[\frac{(x-1) (y-1) Q^2 x y^3 m_c^2 z^5}{(z-1)^2}+\frac{(x-1) Q^2 x y^3 m_b m_c z^5}{(z-1)^2}-\frac{Q^2 (x-1) x y^3 m_b m_c z^4}{(z-1)^2}+\frac{(y-1) y m_b m_c^3 z^3}{(z-1)^2}\right.\right.\\
\nonumber&&+\frac{y m_b^2 m_c^2 z^3}{(z-1)^2}+\frac{(x-1) Q^2 x y^4 m_b m_c z^3}{(y-1)^2}-\frac{Q^2 (x-1) x y^3 m_b m_c z^3}{(y-1)^2}+\frac{(x-1) (z-1) Q^2 x y^3 m_b^2 z^2}{(y-1)^2}+\frac{(y-1) (z-1) Q^2 x m_c^2 z^2}{(x-1)^2}\\
\nonumber&&+\frac{(y-1) (z-1) Q^2 m_b m_c z^2}{(x-1)^2}+\frac{(y-1) (z-1) Q^2 m_b m_c z^2}{x}-\frac{Q^2 x (y-1) (z-1) m_b m_c z^2}{(x-1)^2}-\frac{y m_b^2 m_c^2 z^2}{(z-1)^2}\\
\nonumber&&-\frac{Q^2 (x-1) (y-1) (z-1) m_b^2 z^2}{x^2}+\frac{y^2 m_b^2 m_c^2 z}{(y-1)^2}+\frac{x m_b^2 m_c^2 z}{(x-1)^2}+\frac{(x-1) m_b^3 m_c z}{x^2}-\frac{x m_b m_c^3 z}{(x-1)^2}-\frac{m_b^2 m_c^2 z}{(x-1)^2}-\frac{y m_b^2 m_c^2 z}{(y-1)^2}-\frac{m_b^2 m_c^2 z}{x}\\
\nonumber&&\left.+\frac{(z-1) y m_b^3 m_c}{(y-1)^2}\right] Q^2+\left[-\frac{4 Q^2 (x-1) x (y-1) y^3 m_c^2 z^5}{(z-1)^2}-\frac{5 Q^2 (x-1) x y^3 m_b m_c z^5}{(z-1)^2}+\frac{3 (x-1) Q^2 x y^3 m_b m_c z^4}{(z-1)^2}\right.\\
\nonumber&&+\frac{5 (x-1) Q^2 x y^3 m_b m_c z^3}{(y-1)^2}-\frac{5 Q^2 (x-1) x y^4 m_b m_c z^3}{(y-1)^2}-\frac{(y-1) y m_b m_c^3 z^3}{(z-1)^2}-\frac{2 y m_b^2 m_c^2 z^3}{(z-1)^2}+\frac{4 (x-1) (y-1) (z-1) Q^2 m_b^2 z^2}{x^2}\\
\nonumber&&+\frac{5 (y-1) (z-1) Q^2 x m_b m_c z^2}{(x-1)^2}-\frac{4 Q^2 x (y-1) (z-1) m_c^2 z^2}{(x-1)^2}-\frac{5 Q^2 (y-1) (z-1) m_b m_c z^2}{(x-1)^2}-\frac{4 Q^2 (x-1) x y^3 (z-1) m_b^2 z^2}{(y-1)^2}\\
\nonumber&&-\frac{2 Q^2 (x-1) x y^3 m_b m_c z^2}{(y-1)^2}-\frac{5 Q^2 (y-1) (z-1) m_b m_c z^2}{x}+\frac{x m_b m_c^3 z}{(x-1)^2}+\frac{2 y m_b^2 m_c^2 z}{(y-1)^2}+\frac{2 m_b^2 m_c^2 z}{(x-1)^2}+\frac{2 m_b^2 m_c^2 z}{x}\\
\nonumber&&+\left(\frac{3 (x-1) x z^2 m_b m_c y^4}{(y-1)^2}+\frac{2 (x-1) (z-1) x z m_b^2 y^3}{(y-1)^2}+\frac{2 (x-1) (y-1) x z^4 m_c^2 y^3}{(z-1)^2}+\frac{3 (x-1) x z^4 m_b m_c y^3}{(z-1)^2}\right.\\
\nonumber&&+\frac{2 (x-1) x z m_b m_c y^3}{(y-1)^2}-\frac{3 (x-1) x z^2 m_b m_c y^3}{(y-1)^2}-\frac{(x-1) x z^3 m_b m_c y^3}{(z-1)^2}+\frac{2 (y-1) (z-1) x z m_c^2}{(x-1)^2}+\frac{3 (y-1) (z-1) z m_b m_c}{(x-1)^2}\\
\nonumber&&+\frac{3 (y-1) (z-1) z m_b m_c}{x}-\frac{3 x (y-1) (z-1) z m_b m_c}{(x-1)^2}-\frac{2 (x-1) (y-1) (z-1) z m_b^2}{x^2}-\frac{2 (y-1) (z-1) m_b m_c}{(x-1)^2 y}\\
\nonumber&&\left.-\frac{2 (y-1) (z-1) m_b m_c}{x^2 y}\right) F\left(m_b,m_c,Q^2\right) z+\frac{2 (y-1) (z-1) Q^2 m_b m_c z}{(x-1)^2 y}+\frac{2 (y-1) (z-1) Q^2 m_b m_c z}{x^2 y}-\frac{2 x m_b^2 m_c^2 z}{(x-1)^2}\\
\nonumber&&\left.\left.-\frac{2 y^2 m_b^2 m_c^2 z}{(y-1)^2}-\frac{(x-1) m_b^3 m_c z}{x^2}-\frac{2 y m_b^2 m_c^2}{(y-1)^2}-\frac{y (z-1) m_b^3 m_c}{(y-1)^2}-\frac{2 m_b^2 m_c^2}{(x-1)^2 y}-\frac{2 m_b^2 m_c^2}{x^2 y}\right] \log \left(F\left(m_b,m_c,Q^2\right)\right)\right\}\\
\nonumber&&+\frac{\langle g_{s}^{2}GG \rangle}{128 \pi ^6}\int_{0}^{1}dx\int_{0}^{1}dy\int_{0}^{1}dz\;-\frac{5 c_2}{(x-1) x (y-1) (z-1)} \log \left(F\left(m_b,m_c,Q^2\right)\right)\\
\nonumber&&\left\{(1-x) x (1-y) y (1-z) z^3 \left(\left(4 x^2-4 x-3\right) y^2 z+y (7-4 z)+4 (z-1)\right) F\left(m_b,m_c,Q^2\right){}^2\right.\\
\nonumber&&\left.+F\left(m_b,m_c,Q^2\right) \left(2 z m_b m_c \left(y^2 z \left(x^2 (6 z-7)+x (7-6 z)-z+1\right)-y (z-1) \left(\left(6 x^2-6 x-1\right) z-1\right)\right.\right.\right.\\
\nonumber&&\left.\left.\left.+(x-1) x y^3 z^2-z+1\right)+2 Q^2 (x-1) x (y-1) y (z-1) z^3\right.\right.\\
\nonumber&&\left.\left. \left(\left(6 x^2-6 x-5\right) y^2 z+y (11-6 z)+6 (z-1)\right)\right)+2 Q^2 z m_b m_c \left(y^2 z \left(x^2 (3-2 z)+x (2 z-3)\right.\right.\right.\\
\nonumber&&\left.\left.+z-1\right)+y (z-1) \left(\left(2 x^2-2 x-1\right) z-1\right)-(x-1) x y^3 z^2+z-1\right)-2 m_b^2 m_c^2 (y z-1)\\
\nonumber&&\left.-2 Q^4 (x-1) x (y-1) y (z-1) z^3 \left(\left(x^2-x-1\right) y^2 z-y (z-2)+z-1\right)\right\}\\
\end{eqnarray}
}
The correlation function for current $J_6$ and $j_6$ is shown as:
{\allowdisplaybreaks
\begin{eqnarray}
\nonumber
\Pi_{1-}^{pert}(Q^{2})&=& \frac{1}{64 \pi ^6}\int_{0}^{1}dx\int_{0}^{1}dy\int_{0}^{1}dz\;c_1 y z^2 F\left(m_b,m_c,Q^2\right){}^2 \log \left(F\left(m_b,m_c,Q^2\right)\right) \\
\nonumber&&\left\{z F\left(m_b,m_c,Q^2\right) \left(-(x-1) x y^2 z m_b m_c+(y-1) (z-1) m_b m_c-4 Q^2 (x-1) x (y-1) y^2 (z-1) z^2\right)+3 m_b^2 m_c^2\right.\\
\nonumber&&\left.+3 Q^4 (x-1) x (y-1) y^2 (z-1) z^3\right\}\\
\nonumber\Pi_{1+}^{pert}(Q^{2})&=& \frac{1}{64 \pi ^6}\int_{0}^{1}dx\int_{0}^{1}dy\int_{0}^{1}dz\;c_1 y z^2 F\left(m_b,m_c,Q^2\right){}^2 \log \left(F\left(m_b,m_c,Q^2\right)\right) \\
\nonumber&&\left\{z F\left(m_b,m_c,Q^2\right) \left((x-1) x y^2 z m_b m_c-(y-1) (z-1) m_b m_c-4 Q^2 (x-1) x (y-1) y^2 (z-1) z^2\right)\right.\\
\nonumber&&\left.-3 Q^2 (x-1) x y^2 z^2 m_b m_c+3 Q^2 (y-1) (z-1) z m_b m_c-3 m_b^2 m_c^2+3 Q^4 (x-1) x (y-1) y^2 (z-1) z^3\right\}\\
\nonumber\Pi_{1-}^{GG}(Q^{2})&=& \frac{\langle g_{s}^{2}GG \rangle}{768 \pi ^6}\int_{0}^{1}dx\int_{0}^{1}dy\int_{0}^{1}dz\\
\nonumber&&c_1 \left\{\frac{z}{F\left(m_b,m_c,Q^2\right)}\right. \\
\nonumber&&\left.\left[\frac{(x-1) (y-1) Q^2 x y^3 m_c^2 z^4}{(z-1)^2}-\frac{y m_b^2 m_c^2 z^2}{(z-1)^2}+\frac{(x-1) (z-1) Q^2 x y^3 m_b^2 z}{(y-1)^2}+\frac{y m_b^2 m_c^2 z}{(z-1)^2}+\frac{(y-1) (z-1) Q^2 x m_c^2 z}{(x-1)^2}\right.\right.\\
\nonumber&&\left.-\frac{Q^2 (x-1) (y-1) (z-1) m_b^2 z}{x^2}+\frac{y m_b^2 m_c^2}{(y-1)^2}+\frac{m_b^2 m_c^2}{(x-1)^2}+\frac{m_b^2 m_c^2}{x}-\frac{x m_b^2 m_c^2}{(x-1)^2}-\frac{y^2 m_b^2 m_c^2}{(y-1)^2}\right] Q^2+\left[\frac{(x-1) Q^2 x y^3 m_b m_c z^5}{(z-1)^2}\right.\\
\nonumber&&-\frac{4 Q^2 (x-1) x (y-1) y^3 m_c^2 z^5}{(z-1)^2}-\frac{Q^2 (x-1) x y^3 m_b m_c z^4}{(z-1)^2}+\frac{(y-1) y m_b m_c^3 z^3}{(z-1)^2}+\frac{2 y m_b^2 m_c^2 z^3}{(z-1)^2}+\frac{(x-1) Q^2 x y^4 m_b m_c z^3}{(y-1)^2}\\
\nonumber&&-\frac{Q^2 (x-1) x y^3 m_b m_c z^3}{(y-1)^2}+\frac{4 (x-1) (y-1) (z-1) Q^2 m_b^2 z^2}{x^2}+\frac{(y-1) (z-1) Q^2 m_b m_c z^2}{(x-1)^2}+\frac{(y-1) (z-1) Q^2 m_b m_c z^2}{x}\\
\nonumber&&-\frac{4 Q^2 x (y-1) (z-1) m_c^2 z^2}{(x-1)^2}-\frac{Q^2 x (y-1) (z-1) m_b m_c z^2}{(x-1)^2}-\frac{4 Q^2 (x-1) x y^3 (z-1) m_b^2 z^2}{(y-1)^2}+\frac{2 y^2 m_b^2 m_c^2 z}{(y-1)^2}+\frac{2 x m_b^2 m_c^2 z}{(x-1)^2}\\
\nonumber&&+\frac{(x-1) m_b^3 m_c z}{x^2}-\frac{x m_b m_c^3 z}{(x-1)^2}-\frac{2 m_b^2 m_c^2 z}{(x-1)^2}\\
\nonumber&&+\left(-\frac{3 (x-1) x z^2 y^4}{(y-1)^2}+\frac{(x-1) x z^3 y^3}{(z-1)^2}+\frac{3 (x-1) x z^2 y^3}{(y-1)^2}-\frac{2 (x-1) x z y^3}{(y-1)^2}-\frac{3 (x-1) x z^4 y^3}{(z-1)^2}+\frac{3 (y-1) (z-1) x z}{(x-1)^2}\right.\\
\nonumber&&\left.-\frac{3 (y-1) (z-1) z}{(x-1)^2}-\frac{3 (y-1) (z-1) z}{x}+\frac{2 (y-1) (z-1)}{(x-1)^2 y}+\frac{2 (y-1) (z-1)}{x^2 y}\right) F\left(m_b,m_c,Q^2\right) m_b m_c z\\
\nonumber&&\left.\left.-\frac{2 y m_b^2 m_c^2 z}{(y-1)^2}-\frac{2 m_b^2 m_c^2 z}{x}+\frac{2 y m_b^2 m_c^2}{(y-1)^2}+\frac{2 m_b^2 m_c^2}{(x-1)^2 y}+\frac{2 m_b^2 m_c^2}{x^2 y}+\frac{(z-1) y m_b^3 m_c}{(y-1)^2}\right] \log \left(F\left(m_b,m_c,Q^2\right)\right)\right\}\\
\nonumber&&+\frac{\langle g_{s}^{2}GG \rangle}{64 \pi ^6}\int_{0}^{1}dx\int_{0}^{1}dy\int_{0}^{1}dz\;3 c_2 \log \left(F\left(m_b,m_c,Q^2\right)\right) \\
\nonumber&&\Bigg\{-3 y z^3 \left[\left(2 x^2-2 x+1\right) y^2 z-2 y z+y+2 (z-1)\right] F\left(m_b,m_c,Q^2\right){}^2+F\left(m_b,m_c,Q^2\right) \left[2 z m_b m_c \left(\left(\frac{1}{x-1}+6\right) y z\right.\right.\\
\nonumber&&\left.\left.+\frac{1-y z}{x}+\frac{1}{1-x}+\frac{y^2 z (1-y z)}{(y-1) (z-1)}\right)+2 Q^2 y z^3 \left(\left(6 x^2-6 x+1\right) y^2 z+y (5-6 z)+6 (z-1)\right)\right]-8 Q^2 y z^2 m_b m_c\\
\nonumber&&+\frac{2 m_b^2 m_c^2 (y z-1)}{(x-1) x (y-1) (z-1)}-2 Q^4 y z^3 \left((x-1) x y^2 z-y z+y+z-1\right)\Bigg\}\\
\nonumber\Pi_{1+}^{GG}(Q^{2})&=& \frac{\langle g_{s}^{2}GG \rangle}{768 \pi ^6}\int_{0}^{1}dx\int_{0}^{1}dy\int_{0}^{1}dz\\
\nonumber&& c_1 \Bigg\{\frac{1}{F\left(m_b,m_c,Q^2\right)}\left[\frac{(x-1) (y-1) Q^2 x y^3 m_c^2 z^5}{(z-1)^2}+\frac{(x-1) Q^2 x y^3 m_b m_c z^5}{(z-1)^2}-\frac{Q^2 (x-1) x y^3 m_b m_c z^4}{(z-1)^2}+\frac{(y-1) y m_b m_c^3 z^3}{(z-1)^2}\right.\\
\nonumber&&+\frac{y m_b^2 m_c^2 z^3}{(z-1)^2}+\frac{(x-1) Q^2 x y^4 m_b m_c z^3}{(y-1)^2}-\frac{Q^2 (x-1) x y^3 m_b m_c z^3}{(y-1)^2}+\frac{(x-1) (z-1) Q^2 x y^3 m_b^2 z^2}{(y-1)^2}+\frac{(y-1) (z-1) Q^2 x m_c^2 z^2}{(x-1)^2}\\
\nonumber&&+\frac{(y-1) (z-1) Q^2 m_b m_c z^2}{(x-1)^2}+\frac{(y-1) (z-1) Q^2 m_b m_c z^2}{x}-\frac{Q^2 x (y-1) (z-1) m_b m_c z^2}{(x-1)^2}-\frac{y m_b^2 m_c^2 z^2}{(z-1)^2}\\
\nonumber&&-\frac{Q^2 (x-1) (y-1) (z-1) m_b^2 z^2}{x^2}+\frac{y^2 m_b^2 m_c^2 z}{(y-1)^2}+\frac{x m_b^2 m_c^2 z}{(x-1)^2}+\frac{(x-1) m_b^3 m_c z}{x^2}-\frac{x m_b m_c^3 z}{(x-1)^2}-\frac{m_b^2 m_c^2 z}{(x-1)^2}-\frac{y m_b^2 m_c^2 z}{(y-1)^2}-\frac{m_b^2 m_c^2 z}{x}\\
\nonumber&&\left.+\frac{(z-1) y m_b^3 m_c}{(y-1)^2}\right] Q^2+\left[-\frac{4 Q^2 (x-1) x (y-1) y^3 m_c^2 z^5}{(z-1)^2}-\frac{5 Q^2 (x-1) x y^3 m_b m_c z^5}{(z-1)^2}+\frac{3 (x-1) Q^2 x y^3 m_b m_c z^4}{(z-1)^2}+\frac{2 m_b^2 m_c^2 z}{x}\right.\\
\nonumber&&+\frac{5 (x-1) Q^2 x y^3 m_b m_c z^3}{(y-1)^2}-\frac{5 Q^2 (x-1) x y^4 m_b m_c z^3}{(y-1)^2}-\frac{(y-1) y m_b m_c^3 z^3}{(z-1)^2}-\frac{2 y m_b^2 m_c^2 z^3}{(z-1)^2}+\frac{4 (x-1) (y-1) (z-1) Q^2 m_b^2 z^2}{x^2}\\
\nonumber&&+\frac{5 (y-1) (z-1) Q^2 x m_b m_c z^2}{(x-1)^2}-\frac{4 Q^2 x (y-1) (z-1) m_c^2 z^2}{(x-1)^2}-\frac{5 Q^2 (y-1) (z-1) m_b m_c z^2}{(x-1)^2}-\frac{4 Q^2 (x-1) x y^3 (z-1) m_b^2 z^2}{(y-1)^2}\\
\nonumber&&-\frac{2 Q^2 (x-1) x y^3 m_b m_c z^2}{(y-1)^2}-\frac{5 Q^2 (y-1) (z-1) m_b m_c z^2}{x}+\frac{x m_b m_c^3 z}{(x-1)^2}+\frac{2 y m_b^2 m_c^2 z}{(y-1)^2}+\frac{2 m_b^2 m_c^2 z}{(x-1)^2}+\left(\frac{3 (x-1) x z^2 y^4}{(y-1)^2}\right.\\
\nonumber&&+\frac{3 (x-1) x z^4 y^3}{(z-1)^2}+\frac{2 (x-1) x z y^3}{(y-1)^2}-\frac{3 (x-1) x z^2 y^3}{(y-1)^2}-\frac{(x-1) x z^3 y^3}{(z-1)^2}+\frac{3 (y-1) (z-1) z}{(x-1)^2}+\frac{3 (y-1) (z-1) z}{x}\\
\nonumber&&\left.-\frac{3 x (y-1) (z-1) z}{(x-1)^2}-\frac{2 (y-1) (z-1)}{(x-1)^2 y}-\frac{2 (y-1) (z-1)}{x^2 y}\right) F\left(m_b,m_c,Q^2\right) m_b m_c z+\frac{2 (y-1) (z-1) Q^2 m_b m_c z}{(x-1)^2 y}\\
\nonumber&&+\frac{2 (y-1) (z-1) Q^2 m_b m_c z}{x^2 y}-\frac{2 x m_b^2 m_c^2 z}{(x-1)^2}-\frac{2 y^2 m_b^2 m_c^2 z}{(y-1)^2}-\frac{(x-1) m_b^3 m_c z}{x^2}\\
\nonumber&&\left.-\frac{2 y m_b^2 m_c^2}{(y-1)^2}-\frac{y (z-1) m_b^3 m_c}{(y-1)^2}-\frac{2 m_b^2 m_c^2}{(x-1)^2 y}-\frac{2 m_b^2 m_c^2}{x^2 y}\right] \log \left(F\left(m_b,m_c,Q^2\right)\right)\Bigg\}\\
\nonumber&&+\frac{\langle g_{s}^{2}GG \rangle}{64 \pi ^6}\int_{0}^{1}dx\int_{0}^{1}dy\int_{0}^{1}dz\;\frac{3 c_2}{(x-1) x (y-1) (z-1)} \log \left(F\left(m_b,m_c,Q^2\right)\right)\\
\nonumber&& \left\{3 (x-1) x (y-1) y (z-1) z^3 \left(\left(2 x^2-2 x+1\right) y^2 z-2 y z+y+2 (z-1)\right) F\left(m_b,m_c,Q^2\right){}^2\right.\\
\nonumber&&+F\left(m_b,m_c,Q^2\right) \left(2 z m_b m_c \left(-y^2 z \left(x^2 (6 z-5)+x (5-6 z)+z-1\right)\right.\right.\\
\nonumber&&\left.\left.+y (z-1) \left(6 x^2 z-6 x z+z+1\right)+(x-1) x y^3 z^2-z+1\right)\right.\\
\nonumber&&\left.\left.+2 Q^2 (1-x) x (y-1) y (z-1) z^3 \left(\left(6 x^2-6 x+5\right) y^2 z-6 y z+y+6 (z-1)\right)\right)\right.\\
\nonumber&&\left.+2 Q^2 z m_b m_c \left(y^2 z \left(x^2 (2 z-1)-2 x z+x+z-1\right)\right.\right.\\
\nonumber&&\left.-y (z-1) \left(2 x^2 z-2 x z+z+1\right)-(x-1) x y^3 z^2+z-1\right)-2 m_b^2 m_c^2 (y z-1)\\
\nonumber&&\left.-2 Q^4 (1-x) x (y-1) y (z-1) z^3 \left(\left(x^2-x+1\right) y^2 z-y z+z-1\right)\right\}\\
\end{eqnarray}
}

\end{document}